\documentclass[11pt,reqno]{amsart}      

\usepackage{amssymb,amsthm}             
\usepackage{a4}                         
\usepackage{changebar}

\numberwithin{equation}{section}        
\renewcommand{\Re}{{\mathbb R}}         
\newcommand{\la}{\langle}               
\newcommand{\ra}{\rangle}               
\newcommand{\half}{\frac{1}{2}}         
\newcommand{\eps}{\epsilon}		
\newcommand{\Lip}{\text{\rm Lip}}       
\newcommand{\tens}{\otimes}             
\newcommand{\Id}{\text{Id}}             
\newcommand{\EE}{\mathcal E}            
\newcommand{\FF}{\mathcal F}            
\newcommand{\VV}{\mathcal V}            
\newcommand{\GG}{\mathcal G}		
\newcommand{\OP}{\mathcal O \mathcal P} 

\newcommand{\aM}{{\bar M}}              
\newcommand{\ame}{{\bar g}}             
\newcommand{\aR}{{\bar R}}              

\newcommand{\hGamma}{\hat \Gamma}       
\newcommand{\hDelta}{\hat \Delta}       
\newcommand{\del}{\delta}               
\newcommand{\hR}{\hat R}                
\newcommand{\hRR}{\hat {\mathcal R}}    
\newcommand{\Junk}{\Omega}		

\newcommand{\MGamma}{{}^M\Gamma}
\newcommand{\NGamma}{{}^N\Gamma}

\newcommand{\UU}{{\mathcal U}}          
\newcommand{\CC}{{\mathcal C}}          
\newcommand{\HH}{\mathcal H}            
\newcommand{\WW}{\mathcal W}		
\newcommand{\LL}{\mathcal L}            
\newcommand{\Lie}{{\mathcal L}}         
\newcommand{\tr}{{\text{\rm tr}}}       
\newcommand{\Lapse}{N}                  
\newcommand{\hme}{\hat g}               
\newcommand{\hnabla}{\hat \nabla}       
\newcommand{\Shift}{X}                  
\newcommand{\DD}{\langle D \rangle}     

\theoremstyle{plain}
\newtheorem{thm}{Theorem}[section]

\newtheorem{lemma}[thm]{Lemma}

\newtheorem{definition}[thm]{Definition}

\newtheorem{remark}{Remark}[section]

\title[Elliptic--hyperbolic systems]{Elliptic--hyperbolic systems and the Einstein equations}
\author[L. Andersson]{Lars Andersson$^1$}
\thanks{$^1$Supported in part by the Swedish Natural
Sciences Research Council (SNSRC),  contract no.  R-RA 4873-307 and the NSF,
contract no. DMS-0104402}
\address{Department of Mathematics\\
University of Miami\\
Coral Gables, FL 33124, USA\\
}
\email{larsa\char'100math.miami.edu}

\author[V. Moncrief]{Vincent Moncrief$^2$}
\thanks{$^2$Supported in part by the NSF, contracts no.
PHY-9732629 and PHY-0098084}
\address{Department of Physics\\
Yale University\\
P.O. Box 208120\\
New Haven, CT 06520, USA}
\email{vincent.moncrief\char'100yale.edu}

\date{February 21, 2002}

\allowdisplaybreaks[2]

\begin{document}
\begin{abstract}
The Einstein evolution equations are studied in a gauge given by a
combination of the constant mean curvature and spatial harmonic coordinate
conditions. This leads to a coupled quasilinear elliptic--hyperbolic system of
evolution equations. We prove that the Cauchy problem is locally strongly 
well--posed and that a continuation principle holds. 

For
initial data satisfying the Einstein constraint and gauge conditions, the
solutions to the elliptic--hyperbolic system defined by the gauge fixed
Einstein evolution equations are shown to give vacuum spacetimes.
\end{abstract}
\maketitle


\section{Introduction}
In order to construct solutions to
classical field equations with constraints, such as the Yang--Mills
and Einstein equations, it is often necessary
to rewrite the system, either by extracting a hyperbolic system, or by
performing a gauge fixing. The gauge fixing may result in a hyperbolic
system, as is the case for example using Lorentz gauge for the Yang--Mills
equations, or space--time harmonic coordinates for the Einstein equations.
For discussions of hyperbolicity and gauge choices for the Einstein
equations, see \cite{klainerman:nicolo:review,friedrich:rendall:cauchy}.

On the other hand, there are interesting gauge choices which lead to a
coupled elliptic--hyperbolic system, such as the Coulomb gauge for
Yang--Mills, which was used in the global existence proof of Klainerman and
Machedon \cite{klainerman:machedon:YM}. For the Einstein equations, the 
constant mean curvature gauge leads to an elliptic equation for
the Lapse function, and to an elliptic--hyperbolic system for
the second fundamental form $k_{ij}$, cf. \cite{ChB:york:wellposed}. 

In this paper we introduce and study a gauge condition for the
Einstein equations, which  is a combination
of constant mean curvature gauge and a  spatial harmonic coordinate
condition. This leads to an elliptic--hyperbolic system 
where the hyperbolic part is
a modified version of the Einstein evolution equations, and where
the elliptic part consists of the defining equations for Lapse and Shift.

\subsection{The gauge fixed vacuum Einstein evolution equations}
Let $M$ be a compact, connected, orientable 
$C^{\infty}$ manifold of dimension $n \geq 2$ and let
$\aM = \Re \times M$. We define $t: \aM \to \Re$ by projection on the first
component. We will consider Lorentz metrics $\ame$ on $\aM$ so that the level
sets of $t$, $M_t = \{t\} \times M$ are Cauchy surfaces. When there is no
room for confusion we write simply $M$ instead of $M_t$.

Given $(\aM, \ame)$, let $T$ be a time--like normal to $M_t$,
let the Lapse function $\Lapse$ and Shift
vectorfield $\Shift$ be defined by
$$
\partial_t = \Lapse T + \Shift ,
$$
and assume $\Lapse > 0$ so that $T$ is future directed.

Let $\{e_i\}_{i=1}^n$, 
be a time independent frame on $M$ and let 
$\{e^i\}_{i=1}^n$ be its dual frame. 
Let $e_0 = T$ and let $\{e_{\alpha}\}_{\alpha=0}^n = \{e_0,e_1,\dots,e_n\}$ 
so that $\{e_{\alpha}\}_{\alpha=0}^n$ is an adapted frame on $\aM$ (which is
neither global nor time--independent), with 
dual frame $\{e^{\alpha}\}_{\alpha=0}^n$. In some cases we will use local
coordinates $x^i$, and use the notation $\partial_i = \partial/\partial x^i$,
for the coordinate frame and the corresponding first derivative operators. 
In the following we will use frame indices, unless otherwise specified, and
let 
greek indices  take values in
$0,1,\dots,n$ while lower case latin indices take values in
$1,\dots,n$.

The Lorentz metric $\ame$ is of the form
\begin{equation}\label{eq:n+1metr}
\ame = - \Lapse^2 dt\tens dt + g_{ij} ( e^i + X^i dt) \tens (e^j + X^j dt) ,
\end{equation}
where $g = g_{ij} e^i \tens e^j$ 
is the induced metric on $M$. The second fundamental form 
$k$ of $M$ in $\aM$ is given by
$$
k_{ij} = - \half \Lie_T \ame_{ij} = - \half \Lapse^{-1} (\partial_t g_{ij} -
\Lie_X g_{ij} ) , 
$$
where $\Lie$ denotes the Lie--derivative operator. 
The vacuum Einstein equations
\begin{equation}\label{eq:vacuum}
\aR_{\alpha\beta} = 0
\end{equation}
can be written as a
system of evolution and constraint equations for
$(g,k)$.
The vacuum Einstein evolution equations are
\begin{subequations}\label{eq:evolution}
\begin{align}
\partial_t g_{ij} &= -2 \Lapse k_{ij} + \Lie_{\Shift} g_{ij} , 
\label{eq:evolution-g}\\
\partial_t k_{ij} &= - \nabla_i \nabla_j \Lapse + \Lapse(R_{ij} + \tr k
k_{ij} - 2 k_{im}k^m_{\ j} ) + \Lie_{\Shift} k_{ij} . \label{eq:evolution-k}
\end{align}
\end{subequations}
and the vacuum constraint equations are
\begin{subequations}\label{eq:constraint}
\begin{align}
R - |k|^2 + (\tr k)^2 &= 0 , \label{eq:constraint-ham} \\
\nabla_i \tr k - \nabla^j k_{ij}  &= 0 . \label{eq:constraint-mom}
\end{align}
\end{subequations}
A solution to the Einstein evolution and constraint equations is a curve
$t \mapsto (g,k,\Lapse, \Shift)$ which satisfies
(\ref{eq:evolution},\ref{eq:constraint}). Assuming sufficient regularity,
the spacetime metric $\ame$ given in terms of
$(g,\Lapse, \Shift)$ by (\ref{eq:n+1metr}) solves the vacuum Einstein
equations (\ref{eq:vacuum}) if and only if the corresponding curve
$(g,k,\Lapse, \Shift)$ solves (\ref{eq:evolution},\ref{eq:constraint}).
The system (\ref{eq:evolution},\ref{eq:constraint}) is not hyperbolic, and to 
get a well--posed evolution problem we must modify the system. We will do
this by fixing the gauge. 

Let $\hme$ be a fixed $C^{\infty}$ Riemann metric on $M$ with Levi--Civita 
covariant derivative $\hnabla$ and Christoffel symbol $\hGamma^k_{ij}$.  
Define
the vector field $V^k$ by
\begin{subequations}\label{eq:Vdef}
\begin{align}\label{eq:Vdef-frame}
V^k &= g^{ij} e^k ( \nabla_i e_j - \hnabla_i e_j) , \\
\intertext{or in local coordinates, }
V^k &= g^{ij} (\Gamma^k_{ij} - \hGamma^k_{ij}) \label{eq:Vdef-coord} .
\end{align}
\end{subequations}
Then $-V^k$ is the tension field of the identity map $\Id: (M,g) \to (M,
\hme)$, so that $\Id$  is harmonic exactly when
$V^k = 0$, see \cite{eells:lemaire:report1} for background on harmonic maps.

The constant mean curvature and spatial harmonic coordinates (CMCSH) 
gauge condition is given by the equations
\begin{subequations}\label{eq:gauge}
\begin{align}
\tr_g k &= t  && \text{\rm (constant mean curvature)} \label{eq:gauge-CMC}\\
V^k  &= 0  &&\text{\rm (spatial harmonic
coordinates)} \label{eq:gauge-SH}
\end{align}
\end{subequations}
\begin{remark}\label{rem:gaugesource}
In this paper we will restrict our attention to the homogeneous gauge
conditions given above. However, it seems likely that the ideas presented
here can be generalized to include gauge source functions, with the gauge
conditions (\ref{eq:gauge}) replaced by for example
\begin{subequations}\label{eq:gauge-source}
\begin{align}
\tr_g k &= t + f^0 ,  \\
V^k &= f^k , 
\end{align}
\end{subequations}
where $f^0, f^k$ are a function and a space--like vector field on $M$,
independent of the data.
\end{remark}

Let the second order operator 
$\hDelta_g$ be defined on symmetric 2--tensors by 
\begin{equation}\label{eq:hDeltadef}
\hDelta_g  h_{ij} = \frac{1}{\mu_g} \hnabla_m ( g^{mn} \mu_g \hnabla_n h_{ij} 
) ,
\end{equation}
where $\mu_g =
\sqrt{\det g}$ is the volume element on $(M,g)$.
Using the identity $\hnabla_m ( g^{mn} \mu_g ) = - V^n \mu_g$,  
$\hDelta_g h_{ij}$ may be written in the form 
$$
\hDelta_g h_{ij} = g^{mn} \hnabla_m \hnabla_m h_{ij} - V^m \hnabla_m h_{ij} .
$$
In particular, if the gauge condition $V= 0$ is satisfied, 
$\hDelta_g h_{ij} = g^{mn} \hnabla_m \hnabla_n h_{ij}$.
A computation, cf. section \ref{sec:CMCSH-loc-proof}, shows that
$$
R_{ij} = - \half \hDelta_g g_{ij} + S_{ij} [g,\partial g] + \del_{ij} .
$$
where the symmetric tensor $\del_{ij}$ is defined by
\begin{equation}\label{eq:deltadef}
\del_{ij} = \half (\nabla_i V_j + \nabla_j V_i) ,
\end{equation}
and $S_{ij}[g,\partial g]$ is at most of 
quadratic order in the first derivatives of $g_{ij}$. 
Thus the system $g_{ij} \mapsto R_{ij} - \del_{ij}$ is quasilinear elliptic.

In order to construct solutions to the Cauchy problem for the system
consisting of the Einstein evolution and constraint equations
(\ref{eq:evolution},\ref{eq:constraint}) together with the gauge conditions
(\ref{eq:gauge}), we will consider the following
modified form of the Einstein evolution equations,
\begin{subequations}\label{eq:evol-mod}
\begin{align}
\partial_t g_{ij} &= -2 \Lapse k_{ij} + \Lie_{\Shift} g_{ij} , 
\label{eq:evol-mod-g}\\
\partial_t k_{ij} &= - \nabla_i \nabla_j \Lapse + \Lapse(R_{ij} + \tr k
k_{ij} - 2 k_{im}k^m_{\ j} - \del_{ij} )
+ \Lie_{\Shift} k_{ij} , \label{eq:evol-mod-k}
\end{align}
\end{subequations}
coupled to the elliptic defining equations for $\Lapse,\Shift$, needed to
preserve the imposed gauge conditions, 
\begin{subequations}\label{eq:defineNX}
\begin{align}
- \Delta \Lapse + |k|^2 \Lapse &= 1 ,  \label{eq:defineN} \\
\Delta \Shift^i + R^i_{\ f} \Shift^f - \Lie_{\Shift} V^i 
&= (- 2 \Lapse k^{mn} + 2 \nabla^m \Shift^n )
e^i(\nabla_m e_n - \hnabla_m e_n)  
\nonumber \\
&\quad
+ 2 \nabla^m \Lapse k_m^i - \nabla^i \Lapse k_m^{\ m} . 
\label{eq:defineX}
\end{align}
\end{subequations}
If $\del_{ij} = 0$, in particular if $V^k = 0$,
then (\ref{eq:evol-mod}) coincides with the Einstein
vacuum evolution equations (\ref{eq:evolution}).
The vacuum Einstein evolution equations in CMCSH gauge is the 
coupled system (\ref{eq:evol-mod}--\ref{eq:defineNX}). 
In view of the fact that $g_{ij} \mapsto R_{ij} - \del_{ij}$ is elliptic,
the system (\ref{eq:evol-mod}) is hyperbolic, and the coupled system 
(\ref{eq:evol-mod}--\ref{eq:defineNX}) is elliptic--hyperbolic.

\section{The Cauchy problem for quasi--linear hyperbolic systems}
\label{sec:localclass}
In this section we prove that the Cauchy problem for a class of quasi--linear
hyperbolic evolution equations, which includes coupled
elliptic--hyperbolic systems of the form
(\ref{eq:evol-mod}--\ref{eq:defineNX}), is strongly well-posed. The
techniques used are not new, cf. 
\cite{majda,taylor:NLPDE,daveiga:CPAM,daveiga:ARMA} for
treatments of various aspects of the problem for classical quasi--linear
hyperbolic systems. 
The methods of
Kato \cite{kato:quasilinear} for general quasi--linear evolution equations
can presumably be used to prove the results stated
here. However, in view of the abstract nature of the techniques involved in
the approach of Kato, we have decided to give a 
reasonably complete treatement of the Cauchy 
problem for the class of equations that is of interest from the point of view 
of applications in this paper and its sequel
\cite{andersson:moncrief:3+1}.

Let $\Lambda[g]$ be
the ellipticity constant of $g$, defined
as the least $\Lambda \geq
1$ so that
\begin{equation}\label{eq:gunif}
\Lambda^{-1} g(Y,Y) \leq \hme (Y,Y) \leq \Lambda g(Y,Y) ,
\quad \forall Y \in TM .
\end{equation}
In the following, all norms and function spaces will be defined with respect
to $\hme$.
Let $\ame$ be defined in terms of $g,\Lapse,\Shift$ by
(\ref{eq:n+1metr}). Let  
\begin{equation}\label{eq:Lambda-ame-def}
\Lambda[\ame] = \Lambda[g] + || \Lapse ||_{L^{\infty}} 
+ || \Lapse^{-1}||_{L^{\infty}} 
+ || \Shift ||_{L^{\infty}} .
\end{equation}
For a curve of metrics $t \mapsto \ame(t)$,
it is convenient to define
\begin{equation}\label{eq:Lambdatdef}
\Lambda(T) = \sup_{t \in [0,T]} \Lambda[\ame(t)] .
\end{equation}

We write $D$ for first order spatial derivatives. The action of $D$ on
tensors is defined
using the covariant derivative $\hnabla$ 
w.r.t. $\hme$.
Let $D\ame$ be the first order spatial derivatives of $\ame$. Then 
\begin{align*}
|Dg | + |D\Lapse| + |D\Shift| &\leq C(\Lambda[\ame]) |D\ame| , \\
|D \ame| &\leq C(\Lambda[\ame]) ( |Dg| + |D\Lapse| + |D\Shift|) ,
\end{align*}
where $|\cdot |$ denotes the pointwise norm.
 
We recall some definitions and 
facts from analysis which will be needed in the proof of
local existence. Let $\Delta_{\hme} = \hme^{ij} \hnabla_i \hnabla_j$ be the
Laplace operator defined with respect to the background metric $\hme$, 
acting on functions or
tensors, and let  
$\DD = (1 - \Delta_{\hme})^{1/2}$. Let $W^{s,p}$ denote the Sobolev
spaces and let $H^s = W^{s,2}$. Then $W^{s,p} = \DD^{-s} L^p$, for $s
\in \Re$, $1 < p < \infty$, with norm 
$$
||u||_{W^{s,p}} = || \DD^s u ||_{L^p} .
$$ 
In case $s$ is a
non--negative integer, $W^{s,p}$, $1\leq p \leq \infty$ is the closure of
$C^{\infty}(M)$ w.r.t. the equivalent norm 
$\sum_{|k|\leq s} || D^k u||_{L^p}$.
We will without further notice use the same notation for spaces of tensor
fields as for spaces of functions on $M$.
 
For $I \subset \Re$ an interval, we use the notation 
$F(I;W^{s,p})$ for the space of curves of class $F$ with values in
$W^{s,p}$. Spaces which will be used are $F=C, C^{0,1}, C^j, L^{\infty},
L^1$, $W^{j,1}$, where $C^{0,1}$ denotes the space of continuous 
functions with one (time) derivative in
$L^{\infty}$. 

We use the notation $\OP^s$ for pseudo--differential operators with symbol in 
the Hormander class $S^s_{1,0}$, see Taylor \cite{taylor:NLPDE} for
details. For $P \in \OP^s$, $s \in \Re$, 
\begin{equation}\label{eq:basicopest}
|| Pu ||_{W^{r,p}} \leq C ||u||_{W^{r+s,p}}, \qquad \text{for $r \in \Re$}. 
\end{equation}
In particular, $\DD^s \in \OP^s$ and $D \in \OP^1$. 

The following basic inequalities will be used. 
Assume $1 < p < \infty$.
\begin{enumerate}
\item Product estimate I (Kato and Ponce \cite[Lemma X4]{kato:ponce}, 
\cite[(3.1.59)]{taylor:NLPDE}).
If $s > 0$, $W^{s,p} \cap L^{\infty}$ is an algebra, and the inequality 
\begin{equation}\label{eq:prodineq}
|| uv ||_{W^{s,p}} \leq C ( || u ||_{L^{\infty}} || v ||_{W^{s,p}} +
|| u ||_{W^{s,p}} || v ||_{L^{\infty}} )
\end{equation}
holds. In particular if $s > n/p$, then 
$||uv||_{W^{s,p}} \leq C ||u||_{W^{s,p}} ||v ||_{W^{s,p}}$.
\item Product estimate II
(special case of \cite[Theorem 9.5 3]{palais:foundations}, see also 
\cite[\S 3.5]{taylor:NLPDE}).
Assume $t_i \geq 0$, $i=1,2$, 
some $t_i >0 $. Then for $s \leq  \min(t_1, t_2, t_1 + t_2 - n/p)$, (where
the inequality must be strict if some $t_i = n/p$),  
\begin{equation}\label{eq:prodineq2}
|| uv ||_{W^{s, p}} \leq C || u ||_{W^{t_1,p}}||v||_{W^{t_2,p}} .
\end{equation}
\item Composition estimate (\cite[\S 3.1]{taylor:NLPDE}, see also
\cite[Theorem 1]{sickel}) Let $1 \leq s < \mu$, and suppose 
$F \in C^{\mu}(\Re)$ with $F(0)=0$, where $C^\mu$ denotes the H\"older space. 
Then for $u \in W^{s,p} \cap L^{\infty}$,
$$
|| F(u)||_{W^{s,p}} \leq C ||u||_{W^{s,p}}(1 + || u||_{L^{\infty}}^{\mu-1}) .
$$

\item
Commutator estimate I (\cite[Prop. 3.6.A]{taylor:NLPDE})
Assume $P \in \OP^s$, 
$s > 0$, $\sigma \geq 0$,
then
\begin{equation}\label{eq:commest}
|| [P,u] v ||_{W^{\sigma,p}} \leq C ( || Du ||_{L^{\infty}} 
|| v ||_{W^{s-1+\sigma,p}}
+ || u ||_{W^{s+\sigma,p}}|| v ||_{L^{\infty}} ) .
\end{equation}
\item Commutator estimate II (\cite[(3.6.2)]{taylor:NLPDE})
Assume $P \in \OP^1$.
Then 
$$
|| [P,u]v||_{L^p} \leq C || Du||_{L^{\infty}} ||v||_{L^p} .
$$
\end{enumerate}
Restricting to Sobolev spaces of integer order, the above inequalities can
be proved using the classical methods of calculus. 

We next introduce the class of nonlinear evolution equations which will be
considered. As our application is to the Einstein evolution equations, we
will consider symmetric 2--tensors on $M$ as the unknowns, but it should be
stressed, that the proof generalizes essentially without change to sections
of general vector bundles over $M$.

We will think of the symmetric 2--tensors 
$u_{ij},v_{ij}$ as sections of the vector bundle $Q$ of symmetric
2--tensors over $(M,g)$ with fiber inner product 
$\la u , v\ra$ given by 
$$
\la u , v \ra = u_{ij} v_{kl} \hme^{ik} \hme^{jl}
$$
and the corresponding norm $|u|$ defined by $|u| = \la u , u \ra^{1/2}$.
The natural fiber inner product on derivatives is 
$$
\la \hnabla u , \hnabla v \ra_g = \la \hnabla_m u , \hnabla_n v \ra g^{mn} ,
$$ 
with corresponding norm $| \hnabla u |_g$. The covariant derivative is 
metric, 
$$
Y \la u , v \ra = \la \hnabla_Y u , v \ra + \la u , \hnabla_Y v
\ra ,
$$ 
where $(\hnabla_Y u)_{ij} = Y^m \hnabla_m u_{ij}$, 
and the rough Laplacian
$\hDelta_g$ on $Q$, defined by 
$$
\hDelta_g h_{ij} = \frac{1}{\mu_g} \hnabla_m ( g^{mn} \mu_g  \hnabla_n
 h_{ij} ) ,
$$ 
cf. (\ref{eq:hDeltadef}), 
is self-adjoint with respect to the 
natural $L^2$ inner product on $Q$, 
$$
\int_M \la \hnabla u , \hnabla v \ra_g \mu_g = - \int_M \la \hDelta_g u , v
\ra \mu_g .
$$
The curvature on $Q$ can be computed in terms of the Riemann tensor 
$\hR^i_{\ jkl}$ of $\hnabla$,  
$$
\hnabla_m \hnabla_n h_{ij} - \hnabla_n \hnabla_m h_{ij} = 
- \hR^l_{\ imn} h_{lj} - \hR^l_{\ jmn} h_{il} . 
$$
Let the operator $L$ with coefficients given by $(g,\Lapse,\Shift)$ be
defined by
\begin{equation}\label{eq:Lform}
L[g,\Lapse,\Shift] \begin{pmatrix} u \\ v \end{pmatrix}
 = \begin{pmatrix}
\partial_t u - \Lapse v - \hnabla_{\Shift} u \\
\partial_t v - \Lapse \hDelta_g u 
- \hnabla_{\Shift} v
\end{pmatrix} .
\end{equation}
We will use the notation $\UU = (u,v)$, $\HH^s = H^s\times H^{s-1}$, 
$\WW^{1,\infty} = W^{1,\infty} \times L^{\infty}$. 
Write
$L[\UU]$ for $L$ given by (\ref{eq:Lform}), with
$(g,\Lapse,\Shift) = (g,\Lapse,\Shift)[\UU]$, 
and consider Cauchy problems of the form 
\begin{equation}\label{eq:cauchy-nonlin}
L[\UU]\UU = \FF[\UU] , \qquad \UU \big{|}_{t = 0}  = \UU^0 .
\end{equation}
Define the space $\CC^k_T(\HH^s)$, $1 \leq k \leq \lfloor s \rfloor$,  to be 
$$
\CC_T^k(\HH^s) = \cap_{0 \leq j \leq k-1} C^j([0,T];\HH^{s-j}) .
$$
\begin{definition} \label{def:time}
A number 
$T > 0$
is called a {\bf time of existence} in $\HH^s$ for the Cauchy problem 
(\ref{eq:cauchy-nonlin})
if there is a unique solution 
$\UU \in C([0,T];\HH^s) \cap C^1([0,T];\HH^{s-1})$ to (\ref{eq:cauchy-nonlin}).
The {\bf maximal time of existence} in $\HH^s$ 
for \ref{eq:cauchy-nonlin} is 
$$
T_+ = \sup \{ T : T \text{ is a time of existence for 
(\ref{eq:cauchy-nonlin}}) \} .
$$
The Cauchy problem (\ref{eq:cauchy-nonlin}) is called {\bf strongly locally
well posed} in $\CC^k(\HH^s)$ if the solution map 
$\UU^0 \to \UU$ is continuous as a map 
$$
\HH^s \to \CC_T^k(\HH^s)
$$
for a time of existence $T = T(\UU^0) > 0$, 
which depends continuously on $\UU^0 \in \HH^s$. The continuity of $\UU^0 \to
\UU$ is called {\bf Cauchy stability}. 
\end{definition}
The following definition states the regularity properties of $L,\FF$
which will imply that the Cauchy problem (\ref{eq:cauchy-nonlin}) is strongly
locally well--posed. 
\begin{definition}\label{def:cauchy}
Let $\VV \subset \HH^s$ be connected and open. 
The system $L[\UU]\UU = \FF[\UU]$ is called {\bf quasi--linear hyperbolic} in
$\VV$, if 
the maps 
\begin{subequations}\label{eq:mapdef}
\begin{align}
\UU &\mapsto h[\UU] = (g[\UU],\Lapse[\UU],\Shift[\UU]), & \VV \to H^s \\
\UU &\mapsto \FF[\UU], & \VV \to \HH^s
\end{align}
\end{subequations}
are defined and continuous,  $\ame = \ame[\UU]$ defined
in terms of $(g,\Lapse,\Shift)[\UU]$ satisfies $\Lambda(\ame) < \infty$ for
$\UU \in \VV$, and 
there is a continuous function $C_L =
C_L(\UU^0)$ such that for each $\UU^0 \in \VV$ the following holds.  
\begin{enumerate}
\item \label{point:ball} 
$B_{1/C_L}^s(\UU^0) \subset \VV$, where $B_{1/C_L}^s(\UU^0)$ is the ball in
$\HH^s$ of radius $1/C_L$, centered at $\UU^0$.
\item \label{point:Lip-new}
The maps (\ref{eq:mapdef}) are Lipschitz on $B_{1/C_L}^s(\UU^0)$ with
Lipschitz constant $C_L$ w.r.t. $\HH^r$, $1 \leq r \leq s$, explicitly
\begin{align*}
|| h[\UU_1] - h[\UU_2] ||_{H^r} &\leq C_L || \UU_1 - \UU_2||_{\HH^r} \\
|| \FF[\UU_1] - \FF[\UU_2] ||_{\HH^r} &\leq C_L || \UU_1 - \UU_2 ||_{\HH^r}
\end{align*}
for all $\UU_1, \UU_2 \in B_{1/C_L}^s(\UU^0)$.
\item \label{point:C1-new}
The maps $h, \FF$ have  
Frechet derivatives $Dh, D\FF$ satisfying
\begin{align*}
|| Dh[\UU] \UU' ||_{H^{s-1}} &\leq C_L ||\UU'||_{\HH^{s-1}} \\
|| D\FF[\UU]\UU' ||_{\HH^{s-1}} &\leq C_L ||\UU'||_{\HH^{s-1}}
\end{align*}
for all $\UU \in B^s_{1/C_L}(\UU^0)$.
\item \label{point:C^k-new}
Let $m$ be an integer, $1 \leq m \leq \lfloor s \rfloor - 1$. 
For all integers $j$, $1 \leq j \leq m$, 
$\{\ell_i\}_{i=1}^j$, $\ell_i \geq 1$, 
$\sum_{i=1}^j \ell_i = m$,  the Frechet 
derivatives $D^j h[\UU]$, $D^j \FF[\UU]$ of order $j$
are Lipschitz functions from 
$B_{1/C_L}^s(\UU^0)$ to the spaces of multilinear maps 
\begin{align*}
\bigoplus_{i=1}^j \HH^{s - \ell_i}  &\to H^{s-m}  , \\
\bigoplus_{i=1}^j \HH^{s - \ell_i}  &\to \HH^{s-m}  , 
\end{align*} 
respectively. We call $m$ the {\bf order of regularity} of
(\ref{eq:cauchy-nonlin}).
\end{enumerate}
\end{definition}
\begin{remark}
The order of regularity determines the regularity of the solution
w.r.t. time. For strong local well--posedness, it is sufficient to have order
of regularity $m = 1$. 
\end{remark}
The following is the main result of this section. 
\begin{thm} \label{thm:loc1st} Let $s > n/2+1$. 
Assume that (\ref{eq:cauchy-nonlin}) is
quasi--linear hyperbolic, regular of order $1 \leq m \leq \lfloor s \rfloor -1$, 
in $\VV \subset \HH^s$. Then the following holds. 
\begin{enumerate}
\item \label{point:wellposed} (Strong local well--posedness)
The Cauchy problem 
(\ref{eq:cauchy-nonlin}) is strongly locally well--posed in 
$\CC^{m+1}(\HH^s)$ with time of existence which can be chosen 
depending only on $M, \Lambda[\ame[\UU^0]], 
|| D \ame[\UU^0] ||_{L^{\infty}}$, $|| \UU^0||_{\HH^s}$, $|| \ame[\UU^0]
||_{H^s}$, and 
the constant $C_L = C_L(\UU^0)$ in 
Definition \ref{def:cauchy}.
\item (Continuation) \label{point:continue}
Let $T_+$ be a maximal time of existence for (\ref{eq:cauchy-nonlin}). 
Then either the solution leaves $\VV$ at $T_+$, or $T_+ = \infty$ or 
\begin{equation}\label{eq:diverge}
\limsup_{t \nearrow T_+} \max(\Lambda[\ame], ||D\ame||_{L^{\infty}},
||\partial_t g||_{L^{\infty}}, C_L) =
\infty.
\end{equation}
\end{enumerate}
\end{thm}
The rest of this section is devoted to the proof of Theorem
\ref{thm:loc1st}. In subsection \ref{sec:energy}, the basic energy estimate,
Lemma \ref{lem:high-energy-orig} is proved. Local existence and uniqueness is
proved 
in subsection \ref{sec:locproof}, Cauchy stability 
is proved in subsection
\ref{sec:stable} and the continuation principle, point \ref{point:continue}
is proved in subsection \ref{sec:cont}. 
\subsection{Energy estimates}\label{sec:energy}
In this subsection, fix $T > 0$, $s \geq 1$, let $\aM = [0,T] \times M$,
and assume that  
$g \in L^{\infty}([0,T];W^{1,\infty})\cap C^{0,1}([0,T];L^{\infty})$
and $\Lapse,\Lapse^{-1},\Shift \in L^{\infty}([0,T];W^{1,\infty})$. 
Unless otherwise stated, all constants in this subsection depend only on
$T$ and $\Lambda(T)$. 
For the applications in this paper, there is 
no loss of
generality in assuming, in this subsection, that all fields
are $C^{\infty}$ on $\aM$. 
We will consider the linear system
\begin{equation}\label{eq:cauchy-lin}
L[g,\Lapse,\Shift]\UU = \FF , \qquad \UU\big{|}_{t = 0} = \UU^0 ,
\end{equation}
where $L$ is given by (\ref{eq:Lform}) and 
$$
\UU = \begin{pmatrix} u \\  v \end{pmatrix}, \quad \UU^0 = \begin{pmatrix}
u^0 \\ v^0 \end{pmatrix} , 
\quad \FF = \begin{pmatrix}F_u \\ F_v \end{pmatrix} .
$$
Given $g,\Shift$, let $\rho$ be defined by
\begin{equation}\label{eq:rhodef}
\rho = -\half (\partial_t g - \Lie_X g ) ,
\end{equation}
so that 
$\Lapse^{-1} \rho$ is the second fundamental form of $M$ in $(\aM, \ame)$.
Then under the present assumptions, 
$\rho \in L^{\infty}(\aM)$.

Define the energy $\EE = \EE(t,\UU)$ by
$$
\EE(t,\UU) = \half \int_{M_t} (|u|^2 + |\hnabla u|^2_g + |v|^2)\mu_g ,
$$

\begin{lemma} \label{lem:dtE}
Assume
that $\UU \in L^{\infty}([0,T];\HH^2)\cap C^{0,1}([0,T];\HH^1)$
is a solution to (\ref{eq:cauchy-lin}).
Then
\begin{equation}\label{eq:dtEest}
| \partial_t \EE | \leq C
(\EE^{1/2} || \FF||_{\HH^1} +(1 + ||\rho||_{L^{\infty}}) \EE) ,
\end{equation}
where $C = C(\Lambda[\ame])$.
\end{lemma}
\begin{proof}A computation shows
\begin{align*}
\partial_t \EE &= \int_{M_t} \left (
\la u, \Lapse v + F_u \ra + \la \hnabla_m u ,
\hnabla_n F_u \ra g^{mn} \right . \\
&\qquad \left . +  \la \hnabla_m u , \hRR_{rn} u \ra \Shift^r g^{mn} 
+ \la v, F_v \ra
\right ) \mu_g \\
&\quad + \int_{M_t} \left ( \la \hnabla_i u , \hnabla_j u \ra  \rho^{ij} 
 -\half  ( |u|^2 + |\hnabla u|_g^2 +|v|^2 ) \tr \rho \right ) \mu_g ,
\end{align*}
where the curvature term $\hRR_{rn} u$ is given by 
$$
\hRR_{rn} u_{ij} = 
+ \hR^l_{\ irn} u_{lj} + \hR^l_{\ jrn} u_{il} .
$$
An application of the Schwartz inequality gives the result.
\end{proof}
Using (\ref{eq:dtEest}) to estimate $\partial_t \EE^{1/2}$
and integrating the resulting inequality gives
\begin{equation}\label{eq:EE1/2ineq}
|\EE^{1/2}(T) - \EE^{1/2}(0)| \leq  C \left ( 
|| \FF ||_{L^1([0,T];\HH^1)} + \int_0^T
(1 + || \rho(t)||_{L^{\infty}} )\EE^{1/2}(t) dt \right ) .
\end{equation}
An application of the Gronwall inequality 
gives
\begin{equation}\label{eq:Eineq}
\EE^{1/2}(T)  \leq 
C e^{C\int_0^T||\rho(t)||_{L^{\infty}}dt}
( \EE^{1/2}(0) +|| \FF ||_{L^1([0,T];\HH^1)}) .
\end{equation}
Higher order regularity is proved by estimating $E_s(T) = E_s(T;\UU)$ defined 
by 
\begin{equation}\label{eq:Esdef}
E_s (T) = || \UU||_{L^{\infty}([0,T];\HH^s)} , \qquad s \geq 1 .
\end{equation}
We have 
\begin{equation}\label{eq:Escommut}
E_s(T;\UU) \leq  C E_1(T; \DD^{s-1} \UU) , \qquad s 
\geq 1 .
\end{equation}
The inequality (\ref{eq:Eineq}) gives
\begin{lemma}\label{lem:basic-energy} Assume $\UU \in
L^{\infty}([0,T];\HH^2) \cap C^{0,1}([0,T];\HH^1)$ 
solves (\ref{eq:cauchy-lin}).
Then with $\rho$ given by (\ref{eq:rhodef}),
\begin{equation}\label{eq:dtEcauchy}
E_1(T;\UU)  \leq C e^{C\int_0^T ||\rho(t)||_{L^\infty} dt}
( E_1(0;\UU) + ||\FF||_{L^1([0,T];\HH^1)} ) . 
\end{equation}
\qed
\end{lemma}
In order to derive higher order energy estimates, we 
will apply (\ref{eq:dtEcauchy}) to $\DD^{s-1} \UU$ using the identity
$$
L \DD^{s-1} \UU = [ L , \DD^{s-1} ] \UU + \DD^{s-1} \FF .
$$
In order to do this we must estimate the commutator $[L , \DD^{s-1} ]
\UU$. This is done in the following Lemma. 
\begin{lemma}\label{lem:Lcommut}
Let $s > 1$. 
Assume $(g,\Lapse,\Shift) \in H^s\cap W^{1,\infty}$, $\UU \in \HH^s \cap
\WW^{1,\infty}$.
Then 
\begin{equation}\label{eq:Lcommut}
|| [L , \DD^{s-1} ] \UU ||_{\HH^1} \leq C(\Lambda[\ame]) 
( || \ame ||_{W^{1,\infty}} ||\UU||_{\HH^s} 
+ || \ame ||_{H^s} ||\UU||_{\WW^{1,\infty}} ) .
\end{equation}
\end{lemma}
\begin{proof}
By construction, $[\partial_t , \DD] = 0$, and hence 
$$
\lbrack L , \DD^{s-1} \rbrack \UU  = \begin{pmatrix} \lbrack \DD^{s-1} ,
\Lapse  \rbrack v + \lbrack \DD^{s-1} , \hnabla_{\Shift} \rbrack u \\ 
\lbrack \DD^{s-1} , \Lapse \hDelta_g  \rbrack u  + \lbrack \DD^{s-1} , \hnabla_X \rbrack 
v \end{pmatrix} .
$$
In the following we will write $D$ for a first order operator with smooth
coefficients, such as given by for example $\hnabla \in \OP^1$, so that
$\hnabla_{\Shift} = BD$ with $B$ of order zero. 
Recall that $[ D , \DD^{s-1} ] \in \OP^{s-1}$. In case $s$ is an integer,
then $\hnabla^{s-1}$ may be used instead of $\DD^{s-1}$ in this proof.  
We need to estimate the following quantities: 
\begin{align*}
|| [ \DD^{s-1} , \Lapse] v ||_{H^1} , && || [ \DD^{s-1} , 
\hnabla_{\Shift} ] u
||_{H^1} , \\
|| [ \DD^{s-1} , \Lapse \hDelta_g ] u ||_{L^2} , &&  
 || [ \DD^{s-1} , \hnabla_{\Shift} ] v ||_{L^2} . 
\end{align*}
We will treat each term separately.
For the first term, the commutator estimate gives 
$$
|| [ \DD^{s-1} , \Lapse ] v ||_{H^1} \leq C || D \Lapse ||_{L^{\infty}} 
|| v ||_{H^{s-1}} + ||\Lapse||_{H^s} || v ||_{L^{\infty}} .
$$
The identity 
\begin{equation}\label{eq:BDcomm2}
[BD , \DD^{s-1}] u = B [D , \DD^{s-1}]u + [B ,\DD^{s-1}] D u ,
\end{equation} 
gives using the product and commutator estimates, 
$$
||[\DD^{s-1}, \hnabla_{\Shift} ]u||_{H^1} \leq C (||\Shift ||_{W^{1,\infty}} ||u ||_{H^s} 
+ || \Shift ||_{H^{s}}||Du||_{L^{\infty}} ) . 
$$
This takes care of the second term. 
Next, the identity
$$
[\DD^{s-1} , BD] v
=  \DD^{s-1} [B,D] v + [ \DD^{s-1} , D] B v + [D\DD^{s-1} , B] v 
$$
gives 
$$
|| [\DD^{s-1} , \hnabla_{\Shift} ] v ||_{L^2} \leq C (
||\Shift ||_{W^{1,\infty}} || v ||_{H^{s-1}} 
+ ||\Shift ||_{H^s}|| v ||_{L^{\infty}} ) ,
$$
which takes care of the fourth term. Finally, for the third term, we may in
view of (\ref{eq:hDeltadef}) 
write the second order operator $P = \Lapse \hDelta_g$ in the form 
$$
Pu = A D^2 u + B D u .
$$
With $D = \hnabla$, we have 
\begin{align*}
AD^2 u &= \Lapse g^{mn} \hnabla_m \hnabla_n  u , \\
B D u &= - \Lapse V^l \hnabla_l u = 
- \Lapse g^{mn} (\Gamma_{mn}^l - \hGamma_{mn}^l ) \hnabla_l  u ,
\end{align*}
with $A \in H^s \cap W^{1,\infty}$, $B \in H^{s-1} \cap L^{\infty}$. 

The term  $[\DD^{s-1} , BD] u$ is can be estimated in $L^2$ by expanding the
commutator and using the product estimates to get 
$$
|| [ \DD^{s-1} , BD ] u ||_{L^2} \leq C ( ||B||_{L^{\infty}} || u ||_{H^s} +
|| B ||_{H^{s-1}} || Du ||_{L^{\infty}} ) . 
$$
The identity
\begin{equation}\label{eq:ADcomm}
[AD^2 , \DD^{s-1}] u = 
A [ D^2, \DD^{s-1}]u + [A ,\DD^{s-1} D]D u 
+ (\DD^{s-1} DA)(D u) ,
\end{equation}
together with the commutator and product estimates gives 
$$
|| [AD^2 , \DD^{s-1} ] u ||_{L^2} \leq C (||A||_{W^{1,\infty}} || u||_{H^s}
+ || A ||_{H^s} || Du||_{L^{\infty}} ) .
$$
We now have 
\begin{multline*}
|| [P, \DD^{s-1} ] u ||_{L^2} \leq C \left ( 
(||A||_{W^{1,\infty}}  + ||B||_{L^{\infty}})||u||_{H^s} \right. \\
 \left . + (||A||_{H^s} + ||B||_{H^{s-1}}) ||Du||_{L^{\infty}} .
\right )
\end{multline*}

This gives 
\begin{align*}
|| [\DD^{s-1} , \Lapse \hDelta_g ] u ||_{L^2} &\leq C (\Lambda[\ame])
\left ( 
(||g||_{W^{1,\infty}} + || \Lapse ||_{W^{1,\infty}}) ||u||_{H^s}
\right . \\
&\quad 
\left . + (|| g||_{H^s} + || \Lapse ||_{H^s} ) ||Du||_{L^{\infty}}
\right ) . 
\end{align*}
Collecting the above and using $\Lambda[\ame],||\ame||, ||D\ame||$ 
for the terms involving norms of $g,\Lapse,\Shift$, gives the result. 
\end{proof}
Using the Gronwall inequality, Lemmas \ref{lem:dtE}, \ref{lem:basic-energy}
and  \ref{lem:Lcommut}, gives the following 
higher order energy estimate for the linear system (\ref{eq:cauchy-lin}).
\begin{lemma}\label{lem:high-energy-orig}
Assume $\ame \in L^{\infty}([0,T];H^r\cap W^{1,\infty})$, 
$\partial_t g \in L^{\infty}([0,T];H^{r-1}\cap L^{\infty})$, $r\geq 2$.
Let $\UU = (u,v)$ be a solution to (\ref{eq:cauchy-lin}) satisfying 
$u \in L^{\infty}([0,T];H^r\cap W^{1,\infty})\cap
C^{0,1}([0,T];H^{r-1})$, $v \in L^{\infty}([0,T];H^{r-1}\cap L^{\infty})
\cap C^{0,1}([0,T];H^{r-2})$, $r \geq 2$.
Then, for $1 \leq s \leq r$, 
there is a constant $C = C(T,\Lambda(T))$ so that
\begin{subequations}
\begin{multline}\label{eq:high-energy-base}
E_s(T;\UU) \leq C e^{C \int_0^T ( 
|| \rho ||_{L^{\infty}} +
||D\ame||_{L^{\infty}}) dt}
 ( E_s(0;\UU) + || \FF ||_{L^1([0,T];\HH^s)}  \\
  + \int_0^T ||\ame||_{H^s} || \UU||_{\WW^{1,\infty}} dt
) .
\end{multline}
In particular, if 
$s > n/2 +1$, 
\begin{equation}\label{eq:high-energy-use}
E_s(T;\UU) \leq 
C e^{C \int_0^T (|| \rho ||_{H^{s-1}}  + || \ame ||_{H^s} ) dt}
( E_s(0;\UU) + || \FF ||_{L^1([0,T];\HH^s)} ) .
\end{equation}
\end{subequations}
\qed
\end{lemma}
To prove (\ref{eq:high-energy-use}), note that if $s> n/2+1$, 
the quantities $|| \rho ||_{L^{\infty}}$, 
$||D\ame||_{L^{\infty}}$, and  
$|| \UU||_{\WW^{1,\infty}}$, are dominated by 
$|| \rho ||_{H^{s-1}}$,  $|| \ame ||_{H^s}$, and  $E_s(T;\UU)$.

\subsection{Local existence and uniqueness} \label{sec:locproof}
This subsection is devoted to the proof of local existence and uniqueness
for (\ref{eq:cauchy-nonlin}). 
The proof is an iteration argument, following  
\cite{majda}. 

Note that by Definition \ref{def:cauchy}, $C_L = C_L(\UU^0)$ is continuous in $\UU^0$.
Further,  $\Lambda[\ame[\UU]], || D\ame[\UU]||_{L^{\infty}} $ can be
estimated in terms of $\Lambda[\ame[\UU^0]], || D\ame[\UU^0]
||_{L^{\infty}}$ and $C_L$. In
the rest of this subsection, unless otherwise stated, constants will depend
only on $M, \Lambda[\ame[\UU^0]], 
|| D \ame[\UU^0] ||_{L^{\infty}}$, $|| \UU^0||_{\HH^s}$, $|| \ame[\UU^0]
||_{H^s}$, and the constant $C_L = C_L(\UU^0)$.

Let $R = 1/C_L$. 
Let $\{\UU^0_m\}_{m=0}^{\infty} \subset C^{\infty} \cap B^s_{R/4}(\UU^0)$ 
be a sequence of approximations of $\UU^0$ given by smoothing. 
We will construct a sequence of approximate solutions
$$
\{\UU_m\}_{m=0}^{\infty} \subset C^1([0,T_*];B_R^s(\UU^0)),
$$ 
for some $T_* > 0$, to be chosen, with initial data for $\UU_m$
given by $\UU^0_m$.
  
Set $\UU_0(t) \equiv \UU^0_0$, $\FF_0 \equiv 0$, and let
$L_0 = L[g_0,\Lapse_0,\Shift_0]$ be the time--frozen version of $L$ defined
by setting 
$(g_0(t),\Lapse_0(t), \Shift_0(t)) \equiv 
(g[\UU^0_0],\Lapse[\UU^0_0], \Shift[\UU^0_0])$.

For $m > 0$, let 
\begin{align*}
(g_m, \Lapse_m, \Shift_m) &= (g[\UU_m], \Lapse[\UU_m], \Shift[\UU_m]), \\ 
L_m &= L[\UU_m] , \\
\FF_m &= \FF[\UU_m] ,
\end{align*} 
and for $m\geq 0$ define $\UU_{m+1}$ to be the solution of the linear Cauchy
problem 
\begin{equation}\label{eq:Umseqdef-new}
L_m \UU_{m+1} = \FF_m, \qquad \UU_{m+1} \big{|}_{t=0} = \UU^0_{m+1} .
\end{equation}
The existence of solutions for the Cauchy problem for linear hyperbolic
problems with smooth coefficients is standard,  the proof given in 
\cite[Theorem 3.3]{sogge:lectures} is easily adapted to the present
situation. 
Suppose $\{ \UU_{m'} \}_{m'=0}^{m}$
is a sequence of solutions to (\ref{eq:Umseqdef-new}) with $\UU_{m'}$ taking
values in $B_R^s(\UU^0)$. 
By point \ref{point:C1-new} of Definition
\ref{def:cauchy}, we get 
\begin{equation}\label{eq:dtg}
|| \partial_t g[\UU_m(t)] ||_{H^{s-1}} 
\leq C_L || \partial_t
\UU_m||_{\HH^{s-1}} .
\end{equation}
Let $\rho_m = - \half (\partial_t g_m - \Lie_{\Shift_m} g_m)$. 
From (\ref{eq:Umseqdef-new}) it follows, using the energy estimate,  
that 
$|| \partial_t \UU_m ||_{\HH^{s-1}}$ is bounded by a constant depending on
$C_L$ and hence in view of (\ref{eq:dtg}) we get a bound on
$||\rho_m||_{H^{s-1}} + ||\ame_m||_{H^s}$. 
Given this estimate, it follows 
from Lemma \ref{lem:high-energy-orig},
that there is a constant
$C_R < \infty$, so that as long as $\{\UU_{m'}\}_{m=0}^m$ takes values in 
$B_R^s(\UU^0)$, the energy estimate
\begin{equation}\label{eq:Es-est-use}
E_s(T; \UU_{m+1} ) \leq C_{R} ( E_s(0; \UU_{m+1} ) 
+ || \FF_m||_{L^1([0,T];\HH^s)})
\end{equation}
holds for $\UU_{m+1}$, for $T \leq 1$. The restriction $T\leq 1$ is made so
that $C_R$ does not depend on $T$. 
We can choose the sequence $\{\UU^0_m\}_{m=0}^{\infty}$ such that 
\begin{align}
\UU^0_m &\in B^s_{R/4}(\UU^0)  \text{ for } m \geq 0 \label{eq:u0mR}\\
C_{R}|| \UU_m(0,\cdot) - \UU_{m'}(0,\cdot) ||_{\HH^s} \label{eq:J1umum0}
&< R/4 \text{ for } m,m' \geq 0 .
\end{align}
We will prove, for a $T_* \leq 1$ to be chosen,
convergence for this sequence in $L^{\infty}([0,T_*];\HH^s)\cap
C^{0,1}([0,T_*];\HH^{s-1})$ 
to a limit 
$$
\UU \in C([0,T_*];\HH^s)\cap C^1([0,T_*];\HH^{s-1}) ,
$$ 
by the following steps:
\begin{enumerate}
\item $\{\UU_m\}_{m=0}^{\infty} \subset L^{\infty}([0,T_*];B_R^s(\UU^0))$, by
Lemma \ref{lem:highnorm}
\item $\{\UU_m\}_{m=0}^{\infty}$ is a
Cauchy sequence in 
$L^{\infty}([0,T_*];\HH^1)\cap C^{0,1}([0,T_*];\HH^0)$,
by
Lemma \ref{lem:lownorm}, 
with limit $\UU$ satisfying
$$
(\UU, \partial_t \UU)
\in C_w([0,T_*];\HH^s \times \HH^{s-1}) \cap L^{\infty}([0,T_*];\HH^s \times
\HH^{s-1}) , 
$$
by Lemma \ref{lem:weakconv}. See (\ref{eq:Cwdef}) below for the definition of
$C_w$. 
\item $\UU$ is a solution to (\ref{eq:cauchy-nonlin}).
\item $\UU \in C([0,T_*];\HH^s)\cap C^1([0,T_*];\HH^{s-1})$, Lemma
\ref{lem:CwtoC}.
\end{enumerate}

\begin{lemma}[Boundedness in high norm] \label{lem:highnorm}
There is a time $T_* >0$ 
such that $\{\UU_m\} \subset L^{\infty}([0,T_*];B_R^s(\UU^0))$.
\end{lemma}
\begin{proof}
We must prove that there is a $T_*>0$ so that if
$\UU_m \in L^{\infty}([0,T_*];B_R^s(\UU_0))$,
then $\UU_{m+1} \in L^{\infty}([0,T_*];B_R^s(\UU^0))$. 
In order to do this we consider
$$
L_m (\UU_{m+1} - \UU_0 ) = \FF_m - L_m \UU_0
$$
It follows from
(\ref{eq:Es-est-use}) that
\begin{align*}
|| \UU_{m+1} - \UU_0||_{L^{\infty}([0,T];\HH^s)} &\leq
C_{R} \left ( ||\UU_{m+1}^0 - \UU_0^0 ||_{\HH^s} + \right. \\
&\quad \left.
|| \FF_m ||_{L^1([0,T];\HH^s)} + ||L_m  \UU_0 ||_{L^1([0,T];\HH^s)}
\right)
\end{align*}
and we see using (\ref{eq:J1umum0}) and the fact that
$||f||_{L^1([0,t])} \leq t || f ||_{L^{\infty}([0,t])}$,
that there is a $T_* \leq 1$ so that if $\UU_m \in
L^{\infty}([0,T_*];B_R^s(\UU^0))$, then
\begin{equation}\label{eq:um+1um0est}
|| \UU_{m+1} - \UU_0||_{L^{\infty}([0,T_*];\HH^s)} < R/2 .
\end{equation}
By construction $\UU_0(t) \equiv \UU^0_0$ and hence it follows from
(\ref{eq:u0mR}) that $\UU_{m+1} \in L^{\infty}([0,T_*];B_R^s(\UU^0))$. 
This
completes the proof of Lemma \ref{lem:highnorm}.
\end{proof}
\begin{lemma}[Convergence in low norm]\label{lem:lownorm}
There is a time $T_* > 0$, 
so that $\{\UU_m\}$ is Cauchy in 
$L^{\infty}([0,T_*];\HH^1)\cap C^{0,1}([0,T_*];\HH^0)$.
\end{lemma}
\begin{remark} In order to handle the cases with
$s < 3$, we show in Lemma \ref{lem:lownorm} that $\{\UU_m\}$ is
Cauchy in $L^{\infty}([0,T_*];\HH^1)\cap
C^{0,1}([0,T_*];\HH^0)$, in particular we have control of $\partial_t v_m$ only
in $H^{-1}$.  In case $s \geq 3$, this can be avoided and
$\HH^1, \HH^0$ can in the rest of this section be replaced by $\HH^2,
\HH^1$. In particular if $s \geq 3$, $\partial_t v_m$ is Cauchy in
$L^{\infty}([0,T_*];L^2)$. 
\end{remark}
\begin{proof} Let $T_*$ be as in Lemma \ref{lem:highnorm}.
Let $T \leq T_*$.
We compute
$$
L_m (\UU_{m+1} - \UU_{m'+1} ) = \FF_m - \FF_{m'} - (L_m - L_{m'})\UU_{m'+1}
$$
and hence
\begin{align*}
|| \UU_{m+1} - \UU_{m'+1} ||_{L^{\infty}([0,T];\HH^1)} &\leq
C_{R} \left ( || \UU_{m+1}^0 - \UU_{m'+1}^0 ||_{\HH^1} \right. \\
&\quad
+ || \FF_m - \FF_{m'} ||_{L^1([0,T];\HH^1)} \\
&\quad \left.
+ || (L_m - L_{m'}) \UU_{m'+1} ||_{L^1([0,T];\HH^1)} \right)
\end{align*}
Using the Lipschitz property of the map $\UU \mapsto (g,\Lapse,\Shift,\FF)$
we see that by possibly decreasing $T_*$ 
we get for all $m,m' \geq m_0$,
\begin{align*}
|| \UU_{m+1} - \UU_{m'+1} ||_{L^{\infty}([0,T_*];\HH^1)} &\leq
C_{R}|| \UU_{m+1}^0 - \UU_{m'+1}^0) ||_{\HH^1} \\
&\quad +
\half || \UU_m - \UU_{m'}||_{L^{\infty}([0,T_*];\HH^1)}
\end{align*}

As $\UU_m^0$ is Cauchy in $\HH^1$, we can by
thinning out the sequence $\{\UU_m^0\}$ get
$$
\sum_m C_{R} || \UU_{m+2}^0 - \UU_{m+1}^0 ||_{\HH^1} < R/2
$$
Let $\beta_m = C_{R} || \UU_{m+1}^0 - \UU_{m}^0 ||_{\HH^1}$, $m \geq 0$ and
$a_m = || \UU_m - \UU_{m-1} ||_{L^{\infty}([0,T_*];\HH^1)}$, $m \geq 1$.
Then, 
$$
a_{m+1} \leq \beta_m + \half a_m, \qquad m \geq 1.
$$
Solving the difference equation
$$
a_{m+1} - \half a_m = \beta_m
$$
gives
$$
\sum_{m=1}^{\infty} a_m = 2 \sum_{m=1}^{\infty} \beta_m + 2 a_1
$$
and which shows $\{\UU_m\}$ is Cauchy in $L^{\infty}([0,T_*];\HH^1)$. 

In the equation of motion, which determines $\partial_t \UU_m$, there will
occur terms of the form $g^{ij} \partial_i \partial_j u_m$ and $X^j \partial_j
v_m$. In order to show that $\partial_t \UU_m$ is Cauchy in
$L^{\infty}([0,T_*];\HH^0)$, we need to show that these terms are Cauchy in
$H^{-1}$. To see this, recall that multiplication is continuous 
$H^s \times H^1 \to H^1$ for $s > n/2 + 1$,
cf. product estimate II. In view of the fact that $H^{-1}$ is the dual to
$H^1$, this implies that multiplication defines a continuous map $H^s \times
H^{-1} \to H^{-1}$ for $s > n/2 +1$. 
It is now clear from the mapping properties of $(g,\Lapse,\Shift,\FF)$
and the equation of motion, that $\{\UU_m\}$ is Cauchy in 
$L^{\infty}([0,T_*];\HH^1) \cap C^{0,1}([0,T_*];\HH^0)$.
\end{proof}
The dual space to $H^s$ is $H^{-s}$. Let $(\phi,u)_{-s,s}$ denote the duality
pairing of $H^{-s}$ with $H^s$. Then
$|(\phi,u)_{-s,s}| \leq ||\phi||_{H^{-s}} || u ||_{H^s} $.
Define the space $C_w([0,T];H^s)$ of weakly continuous
functions on $[0,T]$ with values in $H^s$, i.e.
\begin{equation}\label{eq:Cwdef}
C_w([0,T];H^s) = \{ u : (\phi, u)_{-s,s} 
\in C([0,T]), \quad \text{ for all } \phi
\in H^{-s} \}
\end{equation}
\begin{lemma}[Weak convergence]\label{lem:weakconv}
Let $\{ u_m \}_{m=1}^{\infty} \subset C([0,T];H^s)$, $s > 0$
be a bounded sequence and assume $\{u_m\}$ is Cauchy in
$L^{\infty}([0,T];H^{s'})$ for
some $s'$, $0 \leq s' < s$. Then there is a
$u \in C_w([0,T];H^s) \cap L^{\infty}([0,T];H^s)$ so
that for all $\phi \in H^{-s}$, $(\phi, u_m)_{-s,s} \to (\phi,u)_{-s,s}$
uniformly in $L^{\infty}[0,T]$ as $m \to \infty$.
\end{lemma}
\begin{proof}
Let $\phi \in H^{-s}$ be arbitrary and fix $\eps > 0$. Let $C$ be a constant
so that $|| u_m ||_{H^s} \leq C$ for all $m$. Let $u$ be the limit of
$\{u_m\}$ in $L^{\infty}([0,T];H^{s'})$. By the uniform bound on
$||u_m||_{L^{\infty}([0,T];H^s)}$, we find $u \in L^{\infty}([0,T];H^s)$.

Recall $H^{-s'}$ is dense in $H^{-s}$, hence there is a
$\phi' \in H^{-s'}$ so that
$|| \phi - \phi'||_{H^{-s}} < \eps/3C$. Since $u_m \to u$ in $H^{s'}$, we may
choose $m$ large enough so that
$||\phi'||_{H^{-s'}} || u_m - u||_{H^{s'}} < \eps/3$.
Then at $t \in [0,T]$,
\begin{multline*}
|(\phi,u_m)_{-s,s} - (\phi,u)_{-s,s}| \leq
|(\phi - \phi' , u_m)_{-s,s}| \\
+ |(\phi', u_m - u)_{-s',s'}| + 
|(\phi' - \phi, u)_{-s,s}| < \eps
\end{multline*}
As $\eps$ was arbitrary, we find that the sequence $(\phi,u_m)_{-s,s} \in C([0,T])$
converges uniformly to $(\phi,u)_{-s,s} \in C([0,T])$. Hence $u \in
C_w([0,T];H^s)$ as $\phi \in H^{-s}$ was arbitrary.
\end{proof}
By Lemma \ref{lem:weakconv},  $\UU_m \to \UU \in C_w([0,T_*];\HH^s)$. 
Next we prove that $\UU$ solves (\ref{eq:cauchy-nonlin}). It is clear from
the construction that $\UU(0) = \UU^0$, so we need to show that
$L[\UU]\UU -\FF[\UU] =0$.  
Consider the sequence $\{\UU_m\}$ defined by (\ref{eq:Umseqdef-new}). 
We compute
$$
L[\UU]\UU -\FF[\UU] = (L[\UU] - L[\UU_m])\UU + L[\UU_m](\UU - \UU_{m+1}) 
 + (\FF[\UU_m] - \FF[\UU]) .
$$
By Lemma \ref{lem:lownorm}, $\{\UU_m\}$ converges to  $\UU$ in 
$L^{\infty}([0,T_*];\HH^1) \cap C^{0,1}([0,T_*];\HH^0)$. 
By the Lipschitz property of $\UU \mapsto
(g,\Lapse,\Shift,\FF)$, the first and third terms in the right 
hand side tends to zero in $L^{\infty}([0,T_*];\HH^1)$ as
$m \to \infty$. It follows from the definition of $L$ and the 
discussion in the proof of Lemma
\ref{lem:lownorm} that the second term tends to zero in
$L^{\infty}([0,T_*];\HH^0)$. 
However, the left hand side is independent of $m$ and hence
equals zero. 
This proves that $\UU$ is a solution to
(\ref{eq:cauchy-nonlin}).

At this stage we know that $t \mapsto (\UU, \partial_t \UU)$ is weakly
continuous and that $\UU$ is a solution to (\ref{eq:cauchy-nonlin}). 
In order to prove 
$\UU \in C([0,T_*];\HH^s)\cap C^1([0,T_*];\HH^{s-1})$, 
we need the following Lemma. 
\begin{lemma}[Continuity]\label{lem:CwtoC}
Assume $\UU$ is a solution to (\ref{eq:cauchy-nonlin}) with $\FF \in \HH^s$,
and assume $(\UU,\partial_t \UU)$ satisfies
$(\UU,\partial_t \UU) \in C_w([0,T];\HH^s\times \HH^{s-1}) 
\cap L^{\infty}([0,T];\HH^s\times \HH^{s-1})$,
$s > n/2 + 1$.
Then $(\UU,\partial_t \UU) \in C([0,T];\HH^s\times \HH^{s-1})$.
\end{lemma}
\begin{proof}
First recall that continuity on $[0,T]$ is equivalent to  right and left
continuity at each $t \in [0,T]$. Changing the direction of time gives an
equation of the same type, so it is sufficient to prove strong 
right continuity at 
$t \in [0,T]$. By a reparametrization, there is no loss of generality in
assuming $t=0$.   

We know from Lemma \ref{lem:weakconv} that $t \mapsto (\UU,\partial_t \UU)$ 
is weakly continuous. In
order to prove strong continuity, we will use the fact that if
$w_m \to w$ weakly, then  $w_m \to w$ strongly if $||w_m||
\to ||w||$, cf. \cite[\S 12, exercise 3]{rudin:funcan}. 
By fixing 
$(g,\Lapse,\Shift)[\UU(t)]$ in the definition of $\EE$, we may define a norm 
\begin{equation}\label{eq:enorm}
||| \UU'  |||_{s;t}^2 = \EE(t; \DD^{s-1} \UU')
\end{equation}
which for fixed $t$, is an equivalent norm to $||\cdot ||_{\HH^{s}}$.
By (\ref{eq:Escommut}), (\ref{eq:EE1/2ineq}) and Lemma \ref{lem:Lcommut}, 
$$
||| \UU(t) |||_{s;t} \to ||| \UU (0)|||_{s;0} \qquad \text{
as } t \searrow 0 .
$$
By compact imbedding, $\UU \in C([0,T];\HH^{s'})$ for $s' < s$, and hence 
we have $g \in C([0,T];H^{s'})$. From this fact it follows easily that 
\begin{equation}\label{eq:normconv}
||| \UU(t) |||_{s;0} \to ||| \UU (0)|||_{s;0} \qquad \text{
as } t \searrow 0 .
\end{equation}
and hence $\UU$ is right continuous in $\HH^s$ at $t=0$. 
The corresponding property for $\partial_t \UU$ follows from equation
(\ref{eq:cauchy-nonlin}). 
\end{proof}

By Lemma \ref{lem:CwtoC}, we have $\UU \in C([0,T_*];\HH^s) \cap
C^1([0,T_*];\HH^{s-1})$. It follows from (\ref{eq:cauchy-nonlin}) and 
the assumptions on the map $\UU \to (g,\Lapse,\Shift, \FF)$ that 
$\UU \in \CC^{m+1}_{T_*}(\HH^s)$, where $m$ is the order of regularity of
(\ref{eq:cauchy-nonlin}).  
The contraction property used 
in the proof of Lemma \ref{lem:lownorm} shows 
that $\UU$ is the unique solution to
(\ref{eq:cauchy-nonlin}). 

\subsection{Cauchy stability}\label{sec:stable}
In this section we will give a proof of Cauchy stability 
following 
\cite{daveiga:ARMA,daveiga:CPAM}. We will give the proof only for the case 
$s = k$, $k$ integer, $k > n/2+1$, 
and assuming that (\ref{eq:cauchy-nonlin})
is regular of order $m=k-1$. 
The proof is easily adapted to noninteger $s > n/2+1$ and
general $m$.

Let $|| \cdot ||_\ell$ denote 
$|| \cdot ||_{\HH^\ell}$ for $\ell$ integer. 
Introduce the norm
$$
||| \UU(t) |||_{k} = \sum_{0 \leq j \leq k-1} || \partial_t^j
\UU(t)||_{\HH^{k-j}}
$$
Define the spaces 
\begin{align*}
\CC_T(\HH^{k}) &= \cap_{0 \leq j \leq k-1} C^j([0,T];\HH^{k-j}), \\
\LL^1_T(\HH^k) &= \cap_{0 \leq j \leq k-1} W^{j,1}([0,T];\HH^{k-j}) ,
\end{align*}
with norms
\begin{align*}
|||\UU|||_{k,T} &= \sum_{0 \leq j \leq k-1} || \UU ||_{C^j([0,T];\HH^{k-j})} \\
\la\la \UU \ra\ra_{k,T} &= \sum_{0 \leq j \leq k-1} || \UU
||_{W^{j,1}([0,T];\HH^{k-j})}
\end{align*}

Consider the linear problem 
$$
L[h] \UU = \FF , \qquad \UU\big{|}_{t=0} = \UU^0
$$
where we use $h$ to denote the coefficients $g,\Lapse,\Shift$ of $L$. 
We will use the convention that $||| h |||_{\ell,T}$ and $\la\la h
\ra\ra_{\ell,T}$ denotes the norm defined analogously to the above but using
$H^k$ instead of $\HH^k$. 
From
the energy estimate, Lemma \ref{lem:high-energy-orig} 
we get 
for $2 \leq \ell \leq
k$, $\ell$ integer, 
\begin{equation}\label{eq:newenergy}
||| \UU|||_{\ell,T} \leq C(||| h |||_{k,T} )
( || \UU^0||_\ell + |||\FF(0)|||_{\ell-1} + 
\la\la \FF\ra\ra_{\ell,T} ) 
\end{equation}
Using the density of $C^{\infty}$ in Sobolev spaces, the proof of the 
following approximation Lemma straightforward. 
\begin{lemma}[Approximation]
Given any $\UU^0 \in \HH^{\ell}, \FF \in \LL^1_T(\HH^{\ell})$, $\ell \geq 2$,
$\ell$ integer, 
there are for
any $\eps > 0$, $\UU^0_{\eps} \in C^{\infty}, \FF_{\eps} \in C^{\infty}$ 
such that 
\begin{align*}
|| \UU^0 - \UU^0_{\eps} ||_{\ell} &< \eps \\
||| \FF(0) - \FF_{\eps}(0) |||_{\ell - 1} &< \eps \\
\la\la \FF - \FF_{\eps} \ra\ra_{\ell,T} &< \eps
\end{align*}
\qed
\end{lemma} 

\subsubsection{Perturbation estimate for $\ell \leq k-1$}
Consider the linear problems 
\begin{align*}
L[h] \UU &= \FF , \qquad \UU \big{|}_{t=0} = \UU^0 \\
L[h'] \UU' &= \FF' , \qquad \UU' \big{|}_{t=0} = \UU^{\prime 0} \\
L[h] \UU_{\eps} &= \FF_{\eps} , \qquad \UU_{\eps} \big{|}_{t=0} = \UU^0_{\eps}
\end{align*}

In the following write $L = L[h]$, $L' = L[h']$ and assume $h,h' \in
\CC_T(\HH^k)$. In applying the energy estimates in the following we will
let $C$ be a constant depending on $T$ as well as on 
$\Lambda[\ame],|||h|||_{k,T},|||h'|||_{k,T}$. All of these quantities are
under our control, using the apriori estimates. 

We calculate
\begin{align*}
L'(\UU' - \UU_{\eps}) &= \FF' - \FF_{\eps} + (L - L')\UU_{\eps} \\
L(\UU - \UU_{\eps}) = \FF - \FF_{\eps}
\end{align*}
Let 
$$
\Junk_{\ell}(\eps) = ||\UU_{\eps}^0||_{\ell} +
|||\FF_{\eps}(0)|||_{\ell-1} + \la\la \FF_{\eps} \ra\ra_{\ell,T} 
$$
so that by (\ref{eq:newenergy}), for $\ell \leq k$,
$$
||| \UU_{\eps} |||_{\ell,T} \leq C \Junk_{\ell}(\eps) .
$$
Then, for $\ell \leq k-1$, 
$$
\la\la (L - L') \UU_{\eps} \ra\ra_{\ell,T} \leq C ||| h - h'|||_{k-1,T} 
\Junk_{\ell+1}(\eps)
$$
Now we get from (\ref{eq:newenergy}), 
for $\ell \leq k-1$, 
\begin{align*}
|||\UU' - \UU_{\eps}|||_{\ell,T} &\leq C \left\{
||\UU^{\prime 0} - \UU^0_{\eps} ||_{\ell} 
+ ||| \FF'(0) - \FF_{\eps}(0)|||_{\ell-1} \right. \\
&\quad \left. + \la\la \FF' - \FF_{\eps} \ra\ra_{\ell,T} 
+ ||| h - h' |||_{k-1,T}\Junk_{\ell+1}(\eps) \right\}
\end{align*}
It is important to note that this works only for $\ell \leq k-1$. 

The energy estimate (\ref{eq:newenergy}), gives when applied to 
$\UU - \UU_{\eps}$ for $\ell \leq
k-1$, 
$$
||| \UU - \UU_{\eps} |||_{\ell,T} \leq \eps C .
$$
Putting this together gives for $\ell \leq k-1$, 
\begin{equation}\label{eq:UUprime}
\begin{split}
||| \UU - \UU'|||_{\ell,T} &\leq C \left\{ \eps + 
|| \UU^{\prime 0} - \UU^0||_{\ell} 
+ ||| \FF(0) - \FF'(0) |||_{\ell-1} \right. \\
&\quad \left. 
+ \la\la \FF - \FF' \ra\ra_{\ell,T} 
+ ||| h - h' |||_{k-1,T} \Junk_{\ell+1}(\eps) \right\}
\end{split}
\end{equation}

\subsubsection{Perturbation estimate for $\ell = k$}
Let 
$$B = \partial_t - \Shift $$
so that 
$$
L\UU = BU - J\UU
$$
where 
$$
J = \Lapse \begin{pmatrix} 0 & I \\ \Delta & 0 \end{pmatrix}
$$
Let $\delta = B\UU$.
Then $\delta$ solves 
$$
L \delta = \bar\FF, \qquad \delta\big{|}_{t=0} = \bar\UU^0
$$
where 
\begin{align*}
\bar\FF &= [B,J]\UU + B\FF \\
\bar\UU^0 &= (J\UU + \FF)\big{|}_{t=0}
\end{align*}
We have the corresponding primed identities. 
\begin{remark}
For general $s$ we have $\bar\FF$ under control in $\HH^{s-1}$, and
this is also true for $\bar\UU^0$. This observation allows one to generalize
the proof to general $s,m$. 
\end{remark}

From (\ref{eq:UUprime}) applied to $\delta - \delta'$, we get 
\begin{align*}
||| \delta - \delta' |||_{k-1,T} &\leq C \left\{
\eps + || \bar \UU^{\prime 0} - \bar \UU^0 ||_{k-1} 
+ ||| \bar \FF(0) - \bar \FF '(0) |||_{k-2} \right. \\
&\quad \left. 
+ \la\la \bar \FF - \bar \FF' \ra\ra_{k-1,T} 
+ ||| h - h' |||_{k-1,T} \Junk_k(\eps) \right\}
\end{align*}
The terms  $||\bar \UU^{\prime 0} - \bar \UU^0||_{k-1}$,
$ ||| \bar \FF(0) - \bar \FF '(0) |||_{k-2}$, 
$\la\la \bar \FF - \bar \FF' \ra\ra_{k-1,T}$ can be estimated in terms of 
$||\UU^{\prime 0} - \UU^0||_{k}$,
$ ||| \FF(0) -  \FF '(0) |||_{k-1}$, 
$\la\la  \FF - \FF' \ra\ra_{k,T}$.

This gives 
\begin{align*}
||| \delta - \delta' |||_{k-1,T} &\leq C \left\{
\eps + ||  \UU^{\prime 0} -  \UU^0 ||_{k} 
+ |||  \FF(0) -  \FF '(0) |||_{k-1} \right. \\
&\quad \left. 
+ \la\la  \FF -  \FF' \ra\ra_{k,T} 
+ ||| h - h' |||_{k-1,T} \Junk_k(\eps) \right\}
\end{align*}
where now $C$ also depends on $||\UU^0||_k, || \UU^{\prime 0}||_k$.

We now use an elliptic estimate for $\UU - \UU'$. 
A computation shows 
\begin{equation}\label{eq:JUUprime}
J ( \UU - \UU' ) = \delta - \delta ' + \FF' - \FF + (J' - J)\UU'
\end{equation}
From the definition of $J$ and standard elliptic theory we get the estimate 
$$
|| \UU ||_k \leq C ( || J\UU ||_{k-1} + ||\UU||_{k-1} ) 
$$
Hence (\ref{eq:JUUprime}) implies the estimate 
$$
||\UU - \UU' ||_k \leq C ( || \delta - \delta' ||_{k-1} 
+ || \FF' - \FF ||_{k-1} 
+ || h - h'||_{k-1} ||\UU'||_{k} + || \UU - \UU'||_{k-1} ) 
$$
Now calculate
\begin{align*}
\partial_t ( \UU - \UU') &= B\UU - B' \UU' + \Shift\UU - \Shift' \UU' \\
&= \delta - \delta' + \Shift ( \UU - \UU') + (\Shift - \Shift')\UU'
\end{align*}
Iterating this estimate gives together with the above 
\begin{equation}\label{eq:kdiff}
\begin{split}
||| \UU - \UU' |||_{k,T} &\leq C \left\{ \eps 
+ || \UU^0 - \UU^{\prime 0} ||_k + ||| \FF(0) - \FF'(0) |||_{k-1} \right. \\
&\quad \left. 
+ \la\la \FF - \FF'\ra\ra_{k,T} + ||| h - h'|||_{k-1,T} \Junk_k(\eps)
\right\}
\end{split}
\end{equation}

\subsubsection{Application to the nonlinear system}
Consider the Cauchy problems 
\begin{align*}
L[\UU] \UU &= \FF[\UU], \qquad \UU\big{|}_{t=0} = \UU^0 , \\
L[\UU'] \UU' &= \FF[\UU'], \qquad \UU' \big{|}_{t=0} = \UU^{\prime 0} . 
\end{align*}
We wish to estimate $||| \UU - \UU' |||_{k,T}$ in terms of $\UU^0 -
\UU^{\prime 0}$. By the assumptions,
if $\UU, \UU'$ are close, 
there is a constant $C_{\Lip}$ (not to be confused with $C_L$), 
such that 
\begin{align*}
|||\FF[\UU](0) - \FF[\UU'](0)|||_{k-1} &\leq C_{\Lip} || \UU^0 - \UU^{\prime
0} ||_{k-1} , \\
\la\la \FF[\UU] - \FF[\UU'] \ra\ra_{k,T} &\leq 
C_{\Lip} \la\la \UU - \UU'\ra\ra_{k,T}  ,  \\
||| h - h' |||_{k-1,T} &\leq C_{\Lip} ||| \UU - \UU'|||_{k-1,T} .
\end{align*}
By (\ref{eq:kdiff}) we now get an estimate of the form 
\begin{align*}
||| \UU - \UU'|||_{k,T} &\leq C \left\{ \eps + 
|| \UU^0 - \UU^{\prime 0} ||_k \right. \\
&\quad \left. + \la\la \UU - \UU' \ra\ra_{k,T} 
+ ||| \UU - \UU' |||_{k-1,T} \Junk_k(\eps) \right\} .
\end{align*}
The term $\la\la \UU - \UU' \ra\ra_{k,T}$ can be eliminated from the right
hand side by an application of the Gronwall inequality. This gives
\begin{equation}\label{eq:final-k-ineq}
||| \UU - \UU'|||_{k,T} \leq C \left\{ \eps + 
|| \UU^0 - \UU^{\prime 0} ||_k 
+ ||| \UU - \UU' |||_{k-1,T} \Junk_k(\eps) \right\} .  
\end{equation}
Now consider a sequence $\{\UU^0_{\alpha}\} \subset \HH^k$, such that 
$\UU^0_{\alpha} \to \UU^0$ in $\HH^k$ as $\alpha \to \infty$. 
Let $\UU_{\alpha}$ solve 
$$
L[\UU_{\alpha}] \UU_{\alpha} = \FF[\UU_{\alpha}], \qquad
\UU_{\alpha}\big{|}_{t=0} = \UU^0_{\alpha}  .
$$
By compact imbedding we can choose a subsequence so that 
$|||\UU_{\alpha} - \UU |||_{k-1,T} \to 0$. It follows from
(\ref{eq:final-k-ineq}), using the fact that $\eps$ was arbitrary, that 
$$
||| \UU_{\alpha} - \UU|||_{k,T} \to 0, \qquad \text{ as } \alpha \to \infty .
$$
This completes the proof of Cauchy stability and point
\ref{point:wellposed} of Theorem \ref{thm:loc1st} is proved. 
\subsection{Continuation}\label{sec:cont}
It remains to prove the continuation principle, point \ref{point:continue}
of Theorem \ref{thm:loc1st}. 
Suppose for a contradiction, (\ref{eq:diverge})
does not hold and that $T_+ < \infty$ is the maximal time of existence.
Then by the proof of local existence, there is a uniform lower bound for the
time of existence with initial data $\UU(t)$, $t < T_+$. This contradicts
$T_+ < \infty$. The fact that $\UU^0 \to T_+$ is continuous follows from 
Cauchy stability. This completes the
proof of Theorem \ref{thm:loc1st}.

\section{The Cauchy problem
for the modified Einstein evolution equations}
\label{sec:CMCSH-loc-proof}
Let $n \geq 2$ and fix $s > n/2 +1$. The vacuum Einstein equations have
special structure in dimension $2+1$ which we do not make use of, 
but we allow $n=2$ here and in
the following sections for
completeness. 

Let $P$ be the operator defined by 
$$
PY^i = 
\Delta \Shift^i + R^i_{\ f} \Shift^f - \Lie_{\Shift} V^i 
- 2 \nabla^m \Shift^n 
e^i(\nabla_m e_n - \hnabla_m e_n) ,
$$
where $V$ is given by (\ref{eq:Vdef}).
Let the operators $B,E$ be defined by 
\begin{align*}
B f &= - \Delta f + |k|^2 f , \\
E f &=  - 2 \nabla^m f k_m^i + \nabla^i f k_m^{\ m} 
+ 2 f k^{mn} ( \Gamma_{mn}^i - \hGamma_{mn}^i ) .
\end{align*}
With $A$ given by 
\begin{equation}\label{eq:Amatrix}
A = \begin{pmatrix} B & 0 \\ E & P \end{pmatrix} , 
\end{equation}
the defining equation (\ref{eq:defineNX}) can be written in the form 
\begin{equation}\label{eq:Asystem}
A \begin{pmatrix} \Lapse \\ \Shift \end{pmatrix} = 
\begin{pmatrix} 1 \\ 0\end{pmatrix} . 
\end{equation}
The operator $A$ is second order elliptic. 

Define $\VV$
to be the set of symmetric covariant tensors $(g,k) \in H^s\times H^{s-1}$
such that 
\begin{subequations}\label{eq:VVdef}
\begin{align}
&\text{ $g$ is a Riemann metric } &\\
& \text{ The operators $B,P: H^2 \to L^2$,  are isomorphisms
at } (g,k) . &
\end{align} 
\end{subequations}
Suppose $\VV$ is nonempty. For $(g,k) \in \VV$, let $C_{NX} = C_{NX}(g, k)$ 
be a constant so that for functions $u$ and
vectorfields $Y$, 
\begin{subequations}\label{eq:CNX}
\begin{align}
||u||_{L^2} &\leq C_{NX} || B u||_{L^2} ,  \\
||Y||_{L^2} &\leq C_{NX} || PY ||_{L^2} .
\end{align}
\end{subequations}
Let $(g^0, k^0) \in \VV$ be given. 
In this section we prove that the Cauchy problem for the system
(\ref{eq:evol-mod},\ref{eq:constraint},\ref{eq:gauge}), with initial data
\begin{equation}\label{eq:CMCSHdata}
(g,k)\Big{|}_{t = 0} = (g^0, k^0) ,
\end{equation} 
is strongly locally well--posed. We will refer to this problem as the
{\bf CMCSH Cauchy problem} with data $(g^0, k^0)$. 

\begin{thm}
\label{thm:CMCSH-loc} 
The CMCSH Cauchy problem with initial data $(g^0, k^0) \in \VV$ is strongly
locally well--posed in $\CC^k(\HH^s)$, $k = \lfloor s \rfloor$. 
In particular, there is a time of existence $T_* > 0$ so that 
the solution map $(g^0, k^0 ) \to (g,k, \Lapse, \Shift) $ is continuous 
$$
H^s \times H^{s-1} \to \CC^k_{T_*} (H^s\times H^{s-1} \times H^{s+1}\times
H^{s+1} ) 
$$
Here $T_*$ can be
chosen to depend only on $C_{NX} (g^0, k^0)$, $\Lambda[\ame^0]$,
$||\ame^0||_{H^s}$ and $||k^0||_{H^{s-1}}$. 
In particular, $T_*$ can be chosen so that it depends
continuously on $(g^0, k^0) \in H^s \times H^{s-1}$. 

Let $T_+$ be the maximal time of
existence of the solution to the CMCSH Cauchy problem with data $(g^0,
k^0)$. Let $C_{NX} = C_{NX}((g(t),k(t))$ be defined by (\ref{eq:CNX}). Then
either $T_+ = \infty$ or 
$$
\limsup_{t \nearrow T_+} \max(\Lambda[\ame], ||D\ame||_{L^{\infty}},
||k||_{L^{\infty}} , C_{NX}) = \infty
$$
\end{thm}
To prove Theorem \ref{thm:CMCSH-loc} we will show that the CMCSH Cauchy
problem is quasi--linear hyperbolic in $\VV$, regular of order $\lfloor s
\rfloor -1$. The result then follows from
Theorem \ref{thm:loc1st}. 
The following Lemma gives the basic estimates used in the proof of Theorem
\ref{thm:CMCSH-loc}. 
\begin{lemma}\label{lem:second}
The set $\VV$ defined by (\ref{eq:VVdef}) 
is an open subset of $H^s \times H^{s-1}$.
Let $\Lapse,\Shift$ be
the solution to (\ref{eq:defineNX}) w.r.t. $(g,k)$ and let $\ame$ be defined
by (\ref{eq:n+1metr}) in terms of $(g,\Lapse,\Shift)$. 
Let the map 
$\sigma: (g,k) \mapsto (\Lapse,\Shift)$
be defined by solving (\ref{eq:defineNX}).

There is a constant $C_L > 0$ depending only on $C_{NX}, \Lambda[\ame],
||D\ame||_{L^{\infty}}$ such  that the following holds. 
\begin{enumerate}
\item \label{point:Ein-Ball} 
$B_{1/C_L}^s(g^0,k^0) \subset \VV$, where $B_{1/C_L}^s(g^0, k^0)$ is the ball 
in $\HH^s$ of radius $1/C_L$ centered at $(g^0, k^0)$. 
\item \label{point:Ein-C1-new}
The Frechet derivative $D\sigma$ satisfies 
$$
|| D\sigma(g,k) (g', k') ||_{H^s} \leq C_L
   ||(g',k')||_{\HH^{s-1}}
$$
for $(g,k) \in B_{1/C_L}^s(g^0, k^0)$. 
\item \label{point:Ein-Lip} $\sigma$ has the Lipschitz property
$$
|| \sigma(g,k) - \sigma(g',k')||_{H^{r+1}} \leq C_L || (g,k) -
(g',k') ||_{\HH^r} .
$$
for all $(g,k), (g',k') \in B_{1/C_L}^s(g^0,k^0)$, $1 \leq r \leq s$. 
\item \label{point:Ein-Ck-new} 
For integers $m$, $j$, $\ell_i$, such that 
$m = \lfloor s \rfloor -1$, $1 \leq j \leq m$, $1 \leq \ell_i$, 
$\sum_i \ell_i = m$, the Frechet derivative 
$D^j \sigma$ is a Lipschitz map from $B^s_{1/C_L}(g^0, k^0)$ to the space of
multilinear maps 
$$
\bigoplus_{i=1}^j \HH^{s-\ell_i} \to H^{s+1-m}
$$
\end{enumerate}
\end{lemma}
\begin{remark}\label{rem:extra-reg} Note that the Lemma gives an estimate for
$\Lapse,\Shift \in H^{s+1}$. This extra regularity for $\Lapse,\Shift$ will
be important later in the proof of Theorem \ref{thm:CMCSH-loc}.
\end{remark}
\begin{proof} Consider the second order elliptic system for 
$(\Lapse, \Shift)$ given by (\ref{eq:defineNX}). 
In this system the zeroth order coefficients depend on 
the Ricci tensor $R_i^{\ j}$
and derivatives of the
vector field $V^i$ given by (\ref{eq:Vdef}), 
and thus, at first glance, appear to contain second derivatives of
$g_{ij}$, i.e. terms in $H^{s-2}$. 
In order to prove that $(N,X) \in H^{s+1} \times H^{s+1}$ for
$(g,k) \in \VV$, we must prove that these second derivative terms cancel. 

It is clear from (\ref{eq:defineNX}) that it is sufficient to restrict our
attention to the system 
$$
Y^i \mapsto  \Delta Y^i + R^i_{\ f} Y^f - \Lie_{Y} V^i
$$
where $V^i$ is given by (\ref{eq:Vdef}).
We have 
$$
\Delta Y_c + R_c^{\ f} Y_f =
g^{ab} \nabla_a ( \nabla_b Y_c + \nabla_c Y_b
- \nabla_m Y^m g_{bc} ) 
$$
We will compute in local coordinates. 
Writing $\nabla_b Y_c + \nabla_b Y_c = \Lie_Y g_{bc}$ in terms of 
$\partial_i$ gives 
$$
\Lie_Y g_{bc} = Y^f \partial_f g_{bc} + g_{bf} \partial_c Y^f
+ g_{cf} \partial_b Y^f .
$$ 
Using this together with $\nabla_m Y^m = \half g^{mn} \Lie_Y g_{mn}$ shows
after some manipulations that the terms in   
$\Delta Y_c + R_c^{\ f} Y_f$ containing second
derivatives of $g_{ab}$ can be written in the form 
\begin{equation}\label{eq:order2terms}
\half  g^{ab} Y^m  \partial_m (\partial_a g_{bc} + \partial_b g_{ac} -
\partial_c g_{ab}).
\end{equation}
On the other hand, using the explicit form of the 
Christoffel symbol, 
$$
\Gamma^i_{jk} = \half g^{il}(\partial_j g_{kl} + \partial_k g_{jl} -
\partial_l g_{kj})
$$ 
and the form of $V^d$ in local coordinates
(\ref{eq:Vdef-coord}) one sees that (\ref{eq:order2terms}) exactly cancels
the terms in 
$g_{cd} \Lie_Y V^d$ containing second order derivatives in $g_{ij}$.

Let the operators $B,P,A$ be as in (\ref{eq:Amatrix}). 
It follows from the fact that $A$ is lower triangular that if 
$B,P: H^2 \to L^2$ are
isomorphisms, then also $A : H^2 \to L^2$ 
is an isomorphism.

With $u = (\Lapse, \Shift)$, (\ref{eq:Asystem}) and hence
(\ref{eq:defineNX}) 
is of the form 
\begin{equation}\label{eq:Aeq}
A (g,Dg,k) u^l = F^l(g,Dg,k)
\end{equation}
where $F$ is a smooth function of its arguments and 
$A$ is a second order elliptic system of the form 
\begin{equation}\label{eq:Aform}
A(g,Dg,k) u^l= 
g^{mn}\partial_m \partial_n u^l + b^{l\ i}_{\ m}(g,Dg,k) \partial_i u^m
+ c^l_m(g,Dg,k) u^m , 
\end{equation}
where $b,c$ are smooth functions of their arguments. It now follows from 
standard elliptic theory, cf. \cite{taylor:NLPDE} that an inequality of the
form 
\begin{equation}\label{eq:elliptic-ineq}
||u||_{H^{r+2}} \leq C ( || Au||_{H^r} + || u ||_{H^r} ), \qquad 0 \leq r
\leq s-1, 
\end{equation}
holds. Uniqueness together with compact imbedding implies that the lower
order term can be eliminated from this inequality. We sketch the standard
argument for this. 

Suppose that an inequality of the form 
$||u||_{H^{r+2}} \leq C ||Au||_{H^r}$ does not hold. Then there is a
sequence $\{u_i\}_{i=1}^{\infty} $ with $||u_i||_{H^{r+2}} = 1$, and
$||Au_i||_{H^r} \to 0$ as $i \to \infty$. Since $M$ is compact, $H^{r+2}$ is
compactly imbedded in $H^r$, and therefore there is a subsequence $\{ u_j\}$
which converges to some $u_*$ which by construction satisfies $Au_*
= 0$. By (\ref{eq:elliptic-ineq}), in fact $u_* \in H^{r+2}$. The existence
of a solution $u_*$ to $Au=0$ contradicts the assumption that equation  
(\ref{eq:Aeq}) has unique solutions. Thus, an inequality of the
form
\begin{equation}\label{eq:Aineqhomog}
|| u ||_{H^{r+2}} \leq C || Au ||_{H^r} , \qquad 0 \leq r \leq s-1,
\end{equation}
holds. 
It follows from the commutator estimates as in Lemma 
\ref{lem:high-energy-orig}, that the constant $C$ in (\ref{eq:Aineqhomog})
depends only on  $C_L$.

It is convenient to use the notation $h = (g,k)$, $\HH^s = H^s \times
H^{s-1}$. We write $A[h] = A[g,Dg,k]$, $F[h] = F(g,Dg,k)$ etc. 
To prove that $C$ depends continuously on $h \in \HH^s$, let $h'$ be close to 
$h$ in $\HH^s$ and calculate
\begin{align*}
||A[h'] u||_{H^r} &\geq ||A[h] u||_{H^r} - ||(A[h'] - A[h])u||_{H^r} \\
&\geq C^{-1} ||u||_{H^{r+2}} - C_1||h' - h||_{\HH^s} ||u||_{H^{r+2}}
\end{align*}
which by choosing $h'$ sufficiently close to $h$ shows that the estimate 
$|| u ||_{H^{r+2}} \leq C' ||Au||_{H^r}$ holds in a neighborhood of $g$ with a
uniform constant $C'$ arbitrarily close to $C$. 
In particular, for $h'$ close to $h$, $A[h']$ is an isomorphism.
From this follows easily the
first part of the Lemma and 
point \ref{point:Ein-Ball}.

Point \ref{point:Ein-C1-new} follows from elliptic estimates for 
(\ref{eq:Aeq}). 
The Lipschitz property, point \ref{point:Ein-Lip}  
for the map $(g,k) \to (\Lapse, \Shift)$ is a consequence of
\ref{point:Ein-C1-new} and the mean value inequality. 
The proof of point \ref{point:Ein-Ck-new} is straightforward, 
the basic tool being the following
product estimate which is a consequence of the product estimate II. Let
$r > n/2$, 
$\ell_i \geq 1$, $i = 1, \dots, j$, $\sum_i \ell_i = m$, $m \leq r$. 
Then multiplication is a bounded multilinear map 
$$
\bigoplus_{i=1}^j H^{r - \ell_i} \to H^{r-m} .
$$
\end{proof}

\begin{proof}[Proof of Theorem \ref{thm:CMCSH-loc}]
We will show that Theorem
\ref{thm:loc1st} applies to
the modified Einstein evolution equations
(\ref{eq:evol-mod},\ref{eq:defineNX}).
In order to do this we must first write the system
(\ref{eq:evol-mod},\ref{eq:defineNX}) with initial data $(g,k)\big{|}_{t =
t_0} = (g^0, k^0)$, as a quasi--linear hyperbolic system 
(in the sense of Definition \ref{def:cauchy}) of the form
$$
L[\UU] = \FF[\UU], \quad \UU\big{|}_{t=t_0} = \UU^0 .
$$
Let
\begin{subequations}\label{eq:uv-gk}
\begin{align}
u_{ij} &= g_{ij} , \label{eq:uv-g} \\
v_{ij} &= - 2 k_{ij} ,
\end{align}
\end{subequations}
Then
$u_{ij}, v_{ij}$ are symmetric 2--tensors. 
We expand the Lie derivatives 
$\Lie_{\Shift} u_{ij}$, $\Lie_{\Shift} v_{ij}$ in terms of $\hnabla$, 
$$
(\Lie_{\Shift} u)_{ij} = \Shift^m \hnabla_m u_{ij} + u_{lj} \hnabla_i \Shift^l 
+ u_{il} \hnabla_j \Shift^l 
$$
and similarly for $\Lie_{\Shift} v_{ij}$. Similarly 
$\nabla_i \nabla_i \Lapse $ can be
written in terms of $\hnabla$ as 
$$
\nabla_i \nabla_j \Lapse = \hnabla_i \hnabla_j \Lapse 
- \half g^{lm} ( \hnabla_i g_{jm} + \hnabla_j g_{im} - \hnabla_m g_{ji} ) 
\hnabla_l \Lapse .
$$

The Ricci tensor $R_{ij}$ of $g_{ji}$ is quasilinear elliptic, up to a gauge
term. This is seen from the identity 
$$
R_{ij} = - \half \hDelta_g g_{ij} 
+ \del_{ij} + S_{ij}
$$
where $S_{ij}[g, \partial g] $ is given by 
\begin{align*}
S_{ij} &= \half ( g_{li} g^{mn} \hR^l_{\ mjn} 
+ g_{lj} g^{mn} \hR^l_{\ min} ) \\
&\quad + \half 
 g^{mn} g^{ls} \left  \{ \hnabla_j g_{ns} \hnabla_l g_{im} 
+ \hnabla_i g_{ml} \hnabla_s g_{jn} 
- \half \hnabla_j g_{ns} \hnabla_i g_{ml} \right .  \\
&\quad \left . + \hnabla_m g_{il} \hnabla_n g_{js} - \hnabla_m g_{il} \hnabla_s g_{jn}
 \right \} \\
& \quad - \half  V^l \hnabla_l g_{ij} , 
\end{align*}
where 
$$
V^l = g^{mn} ( \Gamma_{mn}^l - \hGamma_{mn}^l ) .
$$
In the computation giving $S_{ij}$, is it convenient to make use of the fact
that $S_{ij}$ is a tensor, and work in a local coordinate system with
$\hGamma^i_{jk} = 0$ at the center of coordinates. 
 
Let $S_{ij} [u, \partial u]$
be the
expression corresponding to $S_{ij} [g, \partial g]$.
Now we may write the system that shows up in the proof of local existence in
the form 
\begin{subequations}\label{eq:vinceform}
\begin{align}
\partial_t u_{ij} &= \Lapse v_{ij} + \Lie_{\Shift} u_{ij} \\
\partial_t v_{ij} &= \Lapse \hDelta_g u_{ij} + \Lie_{\Shift} v_{ij} + F_{ij}
\end{align}
\end{subequations} 
with 
\begin{align}
F_{ij} &= \Lapse \left (  - 2
S_{ij} - \half u^{mn} v_{mn} v_{ij} + v_{im} v_{nj} u^{mn}  \right )
\nonumber \\
&\quad  +  2 \left [   \hnabla_i \hnabla_j \Lapse 
- \half u^{lm} ( \hnabla_i u_{jm} + \hnabla_j u_{im} - \hnabla_m u_{ji} )
\hnabla_l \Lapse \right ] \label{eq:Fvince}
\end{align}
where $u^{mn} = (u^{-1})_{mn}$. 

In order to apply the energy estimates as presented in section
\ref{sec:localclass}, we further expand the Lie derivative in terms of
$\hnabla$. This gives the system in the form 
\begin{subequations}\label{eq:larsform} 
\begin{align}
\partial_t u_{ij} &= \Lapse v_{ij} + \hnabla_{\Shift} u_{ij} + \FF_{1ij} \\
\partial_t v_{ij} &= \Lapse \hDelta_g u_{ij} + \hnabla_{\Shift} v_{ji} +
\FF_{2ij} 
\end{align}
\end{subequations} 
where 
\begin{align*}
\FF_{1ij} &=   u_{lj} \hnabla_i \Shift^l 
+ u_{il} \hnabla_j \Shift^l \\
\FF_{2ij} &=  F_{ij} +  v_{lj} \hnabla_i \Shift^l 
+ v_{il} \hnabla_j \Shift^l 
\end{align*} 
with $F_{ij}$ given by (\ref{eq:Fvince}). 
Finally, assuming that the defining equations 
for $\Lapse, \Shift$ have unique solutions,
let  $\Lapse = \Lapse[u,v]$, $\Shift = \Shift[u,v]$ be given by
(\ref{eq:defineNX}).
Then (\ref{eq:evol-mod}) takes the form
\begin{equation}\label{eq:1storder}
\begin{split}
\partial_t u &= \Lapse v + \hnabla_{\Shift} u +   \FF_1[u,v] \\
\partial_t v &= \Lapse\hDelta_g u + \hnabla_{\Shift} v + \FF_2[u,v]
\end{split}
\end{equation}
Introduce the notation $\UU = (u,v)$,
then $g = g[\UU], k = k[\UU]$ and $\FF[\UU] =
(\FF_1[\UU],\FF_2[\UU])$. Further, via (\ref{eq:defineNX}) we have $\Lapse =
\Lapse[\UU], \Shift = \Shift[\UU]$.

Then we can write equation (\ref{eq:1storder}) in the form
\begin{equation}\label{eq:1stabstract}
L[\UU] \UU = \FF[\UU] , \quad \UU\big{|}_{t=t_0} = \UU^0
\end{equation}
with $L$ given by (\ref{eq:Lform}), and $\UU^0$ given in terms of $(g^0,
k^0)$ by (\ref{eq:uv-gk}). Clearly (\ref{eq:1stabstract}) is a system of the
form considered in Definition \ref{def:cauchy}. 
From the definition of $\FF$ and Lemma \ref{lem:second}, it follows that
the CMCSH Cauchy problem with data in $\VV$ is quasilinear hyperbolic,
regular of order $m = \lfloor s \rfloor -1$, and
hence Theorem \ref{thm:CMCSH-loc} follows from Theorem \ref{thm:loc1st}. 
\end{proof}
\begin{remark}The proof that $\FF : \HH^s \to \HH^s$ is the only place in
the proof of Theorem \ref{thm:CMCSH-loc}
where the ``additional regularity'' $\Lapse,\Shift \in H^{s+1}
\times H^{s+1}$ is used, cf. Remark \ref{rem:extra-reg}.
\end{remark}

\section{Evolution of gauges and constraints}\label{sec:gauge-evol}
Let $n \geq 2$
and fix $s> n/2+1$. Let $\hme$ be a fixed
$C^{\infty}$ metric on $M$ with Levi-Civita covariant derivative $\hnabla$. 
Introduce the constraint and gauge quantities $(A,F,V^i, D^i)$ by
\begin{subequations} \label{eq:defConstr}
\begin{align}
A &= \tr k - t , \\
V^k &= g^{ij} e^k( \nabla_i e_j - \hnabla_i e_j ) 
\label{eq:Vdef-late}\\
F &= R + (\tr k)^2 - |k|^2 - \nabla_i V^i ,\\
D_i &= \nabla_i \tr k - 2\nabla^m k_{mi} .
\end{align}
\end{subequations}
Let also 
\begin{equation}\label{eq:delijdef}
\del_{ij} = \half ( \nabla_i V_j + \nabla_j V_i) ,
\end{equation}
A calculation shows the constraint and gauge quantities $(A,F,V^i,D^i)$
satisfy a hyperbolic system when $(g,k, \Lapse, \Shift)$ solve 
the modified evolution
equation (\ref{eq:evol-mod}--\ref{eq:defineNX}). 
Using an energy estimate for this hyperbolic
system, we will prove that if $(A,F,V^i,D^i) = 0$ for the initial data
$(g^0, k^0)$, then $(A,F,V^i, D^i) \equiv 0$ along the solution curve
$(g,k,\Lapse,\Shift)$ with
\begin{subequations}\label{eq:gkNXreg}
\begin{align}
(g,k) &
\in \cap_{0 \leq \ell \leq m} \leq C^{\ell}((T_0,T_1);H^{s-\ell}
\times H^{s-1-\ell} ), \\
(\Lapse, \Shift) &
\in \cap_{0 \leq \ell \leq m} 
C^{\ell} ((T_0,T_1);H^{s+1-\ell}\times H^{s+1-\ell} ) ,
\end{align}
\end{subequations}
where $m = \lfloor s \rfloor -1$, 
constructed in Theorem \ref{thm:CMCSH-loc}.
In computing the time derivatives of the constraint quantities 
$A,F,V,D$, we
note the fact that the effect of Lie dragging $g_{ij}, k_{ij}$ by
$\Shift$, is that the quantities $A,F,V,D$ are also Lie dragged.
We have using (\ref{eq:evol-mod}--\ref{eq:defineNX}) and 
(\ref{eq:defConstr}),
\begin{subequations}\label{eq:constr-evol}
\begin{align}
\partial_t V^i &= \Lie_{\Shift} V^i + \Lapse D^i
\label{eq:dtV} \\
\partial_t A &=  \Lie_{\Shift} A + \Lapse F\label{eq:dtA} \\
\partial_t F &= \Lie_{\Shift} F + \nabla_i \Lapse D^i + 2 \Lapse F \tr k
+ 2 \Lapse k^{ij} \del_{ij} + V^f \nabla_f (\Lapse \tr k )
\nonumber \\
&\quad
+ \Lapse \Delta A + 2 \nabla_i \Lapse \nabla^i A \\
\partial_t D_i &=  \Lie_{\Shift} D_i + \nabla_i \Lapse F + \Lapse \tr k D_i
+ \Lapse ( \Delta V_i + R_{if} V^f )  \nonumber \\
&\quad + \Lapse \tr k \nabla_i A
+ 2 \nabla^m \Lapse \del_{mi}
\end{align}
\end{subequations}
In doing these computations we have used the expressions
for the Frechet derivatives of of $\Gamma$ and $R$,
\begin{align}
D\Gamma^i_{jk}. h &= \half g^{im} ( \nabla_j h_{km} + \nabla_k h_{jm} - \nabla_m
h_{jk} ) \\
D R . h &= - \Delta  h_i^{\ i} + \nabla^i \nabla^j h_{ij} - R_{ij} h^{ij}
\end{align}
and the identities
\begin{align}
-2 \Delta \nabla_j X^j + \nabla^i \nabla^j (\nabla_i X_j + \nabla_j X_i) &=
\nabla_m R X^m + 2 R^{im} \nabla_i X_m \\
\Lie_Y h_{ij} &= Y^m \nabla_m h_{ij} + h_{mj} \nabla_i Y^m
+ h_{im} \nabla_j Y^m
\end{align}
The
leading order terms in (\ref{eq:constr-evol}) are
\begin{align*}
\partial_t A &\cong \Lapse F \\
\partial_t F &\cong \Lapse \Delta A \\
\partial_t V^i &\cong \Lapse D^i \\
\partial_t D_i &\cong \Lapse (\Delta V_i + R_{if} V^f)
\end{align*}
Using the product and composition 
estimates stated in section \ref{sec:localclass} and the
definition of $R_{ij}$ in terms of $g_{ij}$, one finds that 
at a Riemann metric $g_{ij}$, 
the map $g_{ij} \mapsto R_{ij}$ is smooth and satisfies 
$$
||R_{ij}||_{H^{s-2}} \leq C(\Lambda[g])||g||_{H^s}(1+||g||_{H^s}) ,
$$
for $s > n/2 + 1$.

Define the energy $\EE = \EE_1 + \EE_2$ by
$$
\EE_1 = \half \int (|A|^2 + |\nabla A|^2 + |F|^2 )\mu_g
$$
and
$$
\EE_2 = \half \int (|V|^2 + |\nabla V|^2 + |D|^2 )\mu_g
$$
The following Lemma gives the energy estimate required for proving that the 
Einstein vacuum constraints and the CMCSH gauge is conserved by the modified 
Einstein evolution equations. 
\begin{lemma}\label{lem:EEbound} Let
$(g,k,\Lapse,\Shift)$ 
be a solution to (\ref{eq:evol-mod},\ref{eq:defineNX}), satisfying 
the regularity condition (\ref{eq:gkNXreg}).
There is a constant 
$C = C(g,k,\Lapse)$ 
so that 
$$
| \partial_t \EE | \leq C \EE
$$
\end{lemma}
\begin{proof}
We compute $\partial_t \EE$ using (\ref{eq:constr-evol}).  
Due to covariance, the $\Lie_X$ terms in
(\ref{eq:constr-evol}) can be dropped.
From the assumptions, it follows that $\partial_t g_{ij} \in H^{s-1} \subset
L^{\infty}$. Therefore, we only need to consider the terms in $\partial_t
\EE$ involving $\partial_t A$, $\partial_t F$, $\partial_t V^i$, and
$\partial_t D_i$. 
Using (\ref{eq:constr-evol}) and performing a partial integration, 
it is easy to check that
$$
|\partial_t \EE_1| \leq C \EE .
$$
It remains to consider $\partial_t \EE_2$.
It is straightforward to show $\partial_t \int |V|^2 \mu_g \leq C \EE$.
It remains to consider 
$$
\partial_t \half \int (\nabla_i V^j \nabla_k V^l g^{ik} g_{jl} + D_i D_j
g^{ij} ) \mu_g
$$
The only terms which need to be considered in detail are those where
$\partial_t$ hits $\Gamma^j_{im}$, $V^j$ or $D_i$. The term involving
$\partial_t \Gamma^j_{im}$ is of the form 
\begin{equation}\label{eq:gammaterm}
\int  (\partial_t \Gamma^j_{im}) V^m \nabla_k V^l g^{ik} g_{jl} \mu_g
\end{equation}
The terms involving $\partial_t V^j$, $\partial_t D_i$ yield, after a partial
integration, the expression 
\begin{equation}\label{eq:Rterm}
\int \Lapse R_{if} V^f D^i \mu_g
\end{equation}
From the assumptions, we have $\partial_t \Gamma^j_{im}$ and $R_{if}$
bounded in $H^{s-2}$. Further, $V$ and $D$ are bounded in $H^1$ and $L^2$,
respectively, by $\EE$. The expressions (\ref{eq:gammaterm}) and
(\ref{eq:Rterm}) are of the form 
$\int uvw $
with $u \in H^{s-2}$, $v \in H^1$, $w \in L^2$. 
By the product rule
(\ref{eq:prodineq2}), with $t_1=s-2$, $t_2=1$, $p=2$, we have since $s >
n/2+1$, $||uv||_{L^2} \leq C ||u||_{H^{s-2}} ||v||_{H^1}$.
An application of 
the Cauchy inequality gives the estimate 
$$
|\int uvw \mu_g | \leq C ||u||_{H^{s-2}} || v ||_{H^1} || w ||_{L^2}.
$$
Together with the above this gives 
$|\partial_t \EE_2| \leq C \EE$.   
This completes the proof of Lemma \ref{lem:EEbound}. 
\end{proof}
Lemma \ref{lem:EEbound} and the Gronwall inequality now shows that if
$\EE = 0$ initially, then $\EE = 0$ along the solution curve. We now have 
\begin{thm}\label{thm:constr}
Let $(g^0, k^0) \in \VV$, and assume $(M,g^0, k^0)$ satisfy the Einstein
vacuum constraint equations (\ref{eq:constraint}) and the gauge conditions
(\ref{eq:gauge}), i.e. 
$$
(A,F,V,D) = 0 .
$$ 
Then the space--time metric 
$\ame$ defined 
in terms of the solution $(g,k,\Lapse,\Shift)$ of the
CMCSH Cauchy problem, is a solution of the Einstein vacuum equations.  
\qed
\end{thm}
\section{Isomorphism property}
Let $n \geq 2$ and fix $s > n/2 +1$. In this section, we use local
coordinates unless otherwise stated. Define $\GG$ to be the set of $(g,k) \in
H^s\times H^{s-1}$ such that 
\begin{subequations}\label{eq:GGdef}
\begin{align}
&\text{ $g$ is a Riemann metric } &\\
& \text{ $(A,F,V,D)=0$ at $(g,k)$} &
\end{align} 
\end{subequations}
Thus $(g,k) \in \GG$ precisely when $(g,k)$ satisfies the constraint
equations (\ref{eq:constraint}) and the gauge conditions (\ref{eq:gauge}). 
In this section we will work in the 
CMC time $t = \tr_g k$. 
For initial data $(g^0 , k^0)$ with $t_0 = \tr_{g^0} k^0$, 
let $(T_-, T_+)$,  $T_- < t_0 < T_+$ be a maximal existence interval, defined
by analogy with the notion of maximal existence time. 
\begin{thm}\label{thm:isom}
Assume that $(g^0 , k^0) \in \GG$ and that $\hme$ has negative sectional
curvature. Then the CMCSH Cauchy problem is strongly locally well--posed
in $\CC^k(\HH^s)$, $k = \lfloor s \rfloor$ and the Lorentz metric $\ame$ 
constructed from the solution $(g,\Lapse,\Shift)$ is a vacuum solution of the
Einstein equations. 
Further, the following continuation principle holds. 
Let $t_0 = \tr_{g^0} k^0 < 0$ and let $(T_-, T_+)$, $T_- < t_0 < T_+$,
be a maximal existence
interval for the CMCSH Cauchy problem in CMC time $t = \tr_g k$. 
Then either $(T_-, T_+) = (-\infty, 0)$ or 
$$
\limsup  ( \Lambda[\ame] + 
||D\ame||_{L^{\infty}} + || k ||_{L^{\infty}} ) = \infty
$$
as $t \nearrow T_+$ or as $t \searrow T_-$.
%
\end{thm}

We begin by considering the operator $P$ defined by 
$$
PY^i = 
\Delta Y^i + R^i_{\ f} Y^f - \Lie_{Y} V^i 
- 2 \nabla^m Y^n 
( \Gamma_{mn}^i - \hGamma_{mn}^i ) 
$$

We will need some material concerning vector and tensor fields along maps. 
Consider a map $\phi: (M,g_{ij})  \to (N,h_{\alpha\beta})$. We use latin
indices for coordinates on $M$ and greek indices for coordinates on $N$. 
The bundles of tensor fields along $\phi$, 
$\tens^k T^* M \tens \phi^{-1} TN$
have natural connections which we denote by $D$. 
Here $\phi^{-1} TN$ is the pullback of
$TN$ to $M$ along $\phi$.   
We work this out in local coordinates
for $k=0,1,2$. 
\begin{align}
D_i v^{\gamma} &= \partial_i v^{\gamma} 
+ \NGamma^{\gamma}_{\alpha\beta} v^{\alpha}\partial_i \phi^{\beta} \\
D_i \xi^{\gamma}_j &= \partial_i \xi^{\gamma}_j - \MGamma_{ij}^k
\xi^{\gamma}_k + \NGamma^{\gamma}_{\alpha\beta} \xi^{\alpha}_j \partial_i
\phi^{\beta} \\
D_i \xi^{\gamma}_{jk} &= \partial_i \xi^{\gamma}_{jk} - 
\MGamma^{\ell}_{ij}\xi^{\gamma}_{\ell k} - \MGamma^{\ell}_{ik}
\xi^{\gamma}_{j\ell}
+ \NGamma^{\gamma}_{\alpha\beta} \xi^{\alpha}_{jk} \partial_i \phi^{\beta}
\end{align}
where if the left hand side is evaluated at $x \in M$, all objects on 
$N$ are evaluated at $\phi(x)$. 
 
On the bundles $\tens^k T^* M \tens \phi^{-1} TN$, there is a natural inner
product $\la \cdot , \cdot \ra$ which is local coordinates is given by 
\begin{align*}
\la v , w \ra &= v^{\alpha} w^{\beta} h_{\alpha\beta} \\
\la \xi , \eta \ra &= \xi^{\alpha}_i \eta^{\beta}_j h_{\alpha\beta}g^{ij}
\end{align*}
and similarly for higher order tensors. It is straightforward to check that
$D$ is metric w.r.t. $\la \cdot , \cdot \ra$. Further, letting 
$\Delta_D v^{\alpha} = g^{ij} D_i D_j v^\alpha$ the identity 
$$
\int_M \la \Delta_D v , w \ra \mu_g = - \int_M \la Dv , Dw \ra \mu_g
$$
holds, 
and thus the operator $\Delta_D$ is self--adjoint with respect to the $L^2$
pairing 
$$
 \int_M \la v , w \ra \mu_g
$$

\begin{lemma}\label{lem:isomorph}
Assume $\hme$ has negative sectional curvature and $V = 0$. Then
$P : H^{s+1} \to H^{s-1}$ is an isomorphism. 
\end{lemma}
\begin{proof}
Let $\phi: M \to M$ be a diffeomorphism. 
Then 
\begin{equation}\label{eq:Vpullback}
(\phi^*V)^i = (\phi^* g)^{mn}\left ( {}^{(\phi^* g)}\Gamma_{mn}^i -
{}^{(\phi^* \hme)}\Gamma_{mn}^i\right  )
\end{equation}
This follows from that fact that the difference of Christoffel symbols
transforms as a tensor, or by a direct computation using the identities
\begin{align*}
{}^{(\phi^* g)} \Gamma_{ij}^k &= \partial_m (\phi^{-1})^k \left (
\partial_i \partial_j
\phi^m + {}^{(g)}\Gamma_{rs}^m \partial_i \phi^r \partial_j \phi^s \right ) \\
(\phi^*V)^i (x) &= \partial_m (\phi^{-1})^i V^m (\phi(x))
\end{align*}
Let $Y$ be a vector field on $M$ and let $\phi_s$ defined by $\partial_s
\phi_s = Y \circ \phi_s$, $\phi_0 = \Id$, be the flow of $Y$.  
A computation shows 
$$
\partial_s \big{|}_{s=0} {}^{(\phi_s^* g)}\Gamma^k_{ij} = 
\half g^{kl}(R_{iljm}Y^m + R_{jlim} Y^m + \nabla_i \nabla_j Y_l + \nabla_j
\nabla_i Y_l ) 
$$
and similarly for ${}^{(\phi_s^*\hme)}\Gamma^k_{ij}$. Recall that 
$$
\partial_s\big{|}_{s=0} \phi^*_s V^i = [Y,V]^i = - \Lie_V Y^i .
$$
Now consider the identity (\ref{eq:Vpullback}) with $\phi$ replaced by
$\phi_s$. Differentiating with respect to $s$ and evaluating at $s=0$, 
gives the identity 
\begin{equation}\label{eq:Pident}
PY^i = g^{mn}(\hnabla_m\hnabla_n Y^i + \hR^i_{\ mjn}Y^j)
\end{equation}
where the indices on the Riemann tensor $\hR$ of $\hme$ are raised and
lowered with $\hme$. Using the definition of $\hnabla$ in terms of $\hGamma$
we have 
\begin{align*}
g^{mn} \hnabla_m \hnabla_n Y^i &= g^{mn} ( \partial_m \hnabla_n Y^i -
\hGamma_{mn}^r \hnabla_r Y^i + \hGamma_{jm}^i\hnabla_n Y^j ) \\
&= g^{mn} (\partial_m \hnabla_n Y^i - \Gamma_{mn}^r \hnabla_r Y^i 
+ \hGamma_{jm}^i\hnabla_n Y^j) 
+ V^r \hnabla_r Y^i 
\end{align*}
where we used the definition of $V$ to replace $\hGamma$ by $\Gamma$.

Now we will think of $Y$ as a vector field along the map $\Id: (M,g) \to
(M,\hme)$. For clarity we will use lower case greek indices for objects
associated with $\hme$. 
Recalling the definition of the connection $D$ and
the operator $\Delta_D$, we see that 
$$
g^{mn} \hnabla_m \hnabla_n Y^{\alpha} = \Delta_D Y^{\alpha} + V^r D_r
Y^{\alpha}  
$$
Hence equation (\ref{eq:Pident}) takes the form 
$$
P Y^{\alpha} = \Delta_D Y^{\alpha} + g^{mn} \hR^{\alpha}_{m\beta n}Y^\beta 
+ V^r D_r Y^{\alpha} 
$$
Now assume as in the statement of the Lemma that $V = 0$. Then 
\begin{equation}\label{eq:Pquadr}
\int \la P Y, Y \ra \mu_g  = - \int_M \la DY, DY \ra \mu_g + 
\int_M g^{mn} \hR_{\alpha m \beta n}Y^{\alpha}
Y^{\beta} \mu_g
\end{equation}
The quadratic form $(Y,W) \to g^{mn} \hR_{\alpha m \beta n} Y^{\alpha}
W^{\beta}$ is symmetric and when $\hme$ has negative sectional curvatures
then there is a $\lambda > 0$ so that 
$$
g^{mn} \hR_{\alpha m \beta n} Y^{\alpha} Y^{\beta} \leq - \lambda^2 |Y|^2
$$
It follows that an estimate of the form 
$$
|| Y ||_{L^2} \leq C(\Lambda[g]) ||PY||_{L^2}
$$
holds, and the Lemma now follows from the fact that $P$ is second order
elliptic of the form (\ref{eq:Aform}), cf. the proof of Lemma
\ref{lem:second}. 
\end{proof}

\begin{proof}[Proof of Theorem \ref{thm:isom}]
Let $P$ be as above, let the operators $B,E,A$ be defined as in section 
\ref{sec:CMCSH-loc-proof}.
By the proof of Lemma \ref{lem:second},
$A$ is an elliptic second order operator of the form (\ref{eq:Aform}).  
Therefore, in
order to prove that $A$ is an isomorphism, 
it is enough to prove that $\ker A = \ker A^* = \{0\}$. 
It follows from the maximum principle that $B$ is an isomorphism as long as
$\tr_g k \ne 0$. 
By Lemma \ref{lem:isomorph}, $P$ is an isomorphism. 
In view of the
fact that $A$ is lower triangular, the isomorphism property now follows from 
the isomorphism property for $B$ and $P$. 

We have now proved that $\GG \subset \VV$ and hence the conclusion of
Theorems \ref{thm:CMCSH-loc} and \ref{thm:constr} apply. It remains to prove
the continuation principle. In view of Theorem
\ref{thm:CMCSH-loc}, it is enough to estimate $C_{NX}$ in terms of
$\Lambda[\ame]$ and $(\tr_g k)^{-1}$. By (\ref{eq:Pquadr}) it follows that 
$||Y||_{L^2} \leq C(\Lambda(g)) ||PY||_{L^2}$. Similarly, we have 
$$
(Bf, f)_{L^2} \geq \int_M |\nabla f |^2 + |k|^2 f^2 \geq C(\Lambda(g))
\frac{(\tr_g
k)^2}{3} ||f ||_{L^2}^2
$$
which gives the corresponding estimate for $B$. 
Now we have bounded $C_{NX}$ in terms of $\Lambda[\ame]$ and $(\tr_g
k)^{-1}$.
Finally, to see that $T_+ \leq 0$, note that as $M$ admits a metric $\hme$
with negative sectional curvature, it admits no metric with nonnegative
scalar curvature, cf. \cite[Corollary A, p. 94]{gromow:lawson:dirac}. In case 
$n=2$, this is a consequence of the uniformization theorem. 
Now suppose that $T_+ > 0$. Then $\tr_g k = 0$ 
must occur during the
evolution, at which instant the scalar curvature of $g$ is nonnegative by the
Hamiltonian constraint (\ref{eq:constraint-ham}), which gives a
contradiction. Therefore we have $T_+ \leq 0$. This completes the proof of
Theorem \ref{thm:isom}. 
\end{proof}
\noindent{\bf Acknowledgements.}
Part of this work was done during a visits to 
the Institute of Theoretical Physics, Santa Barbara, Universit\'e Paris VI,
the Albert Einstein Institute, Golm, 
and L'Institut des Hautes \'Etudes Scientifiques, Paris. 
The paper
was finished while the authors were 
enjoying the hospitality
and support of the Erwin Schr\"odinger Institute, Vienna.





\providecommand{\bysame}{\leavevmode\hbox to3em{\hrulefill}\thinspace}

\end{document}